\documentclass[12pt]{article}

\usepackage{bbm}
\usepackage{epsfig}
\usepackage{array}
\usepackage{float}
\usepackage{dsfont}
\usepackage{amstext}
\usepackage{rotating}
\usepackage{a4}
\usepackage{a4wide}
\usepackage{cite}

\def\be{\begin{equation}}
\def\ee{\end{equation}}
\def\gs{\mathrel{
   \rlap{\raise 0.511ex \hbox{$>$}}{\lower 0.511ex \hbox{$\sim$}}}}
\def\ls{\mathrel{
   \rlap{\raise 0.511ex \hbox{$<$}}{\lower 0.511ex \hbox{$\sim$}}}}

\newcommand{\ba}{\begin{array}{c}}
\newcommand{\baz}{\begin{array}{cc}}
\newcommand{\bad}{\begin{array}{ccc}}
\newcommand{\bav}{\begin{array}{cccc}}
\newcommand{\baf}{\begin{array}{ccccc}}
\newcommand{\bea}{\begin{equation} \begin{array}{c}}
\newcommand{\eea}{ \end{array} \end{equation}}
\newcommand{\ea}{\end{array}}
\newcommand{\D}{\displaystyle}

\begin{document}

\title{
\hfill {\small HRI-P-08-03-001}\\
\vglue -0.3cm
\hfill {\small arXiv: 0803.0423 [hep-ph]} 
\vskip 0.5cm
\Large \bf
On Probing $\theta_{23}$ in Neutrino Telescopes } 
\author{
Sandhya Choubey$^a$\thanks{email: \tt sandhya@hri.res.in}~~,
~~Viviana Niro$^b$\thanks{email: 
\tt viviana.niro@mpi-hd.mpg.de}~~,
~~Werner Rodejohann$^b$\thanks{email: 
\tt werner.rodejohann@mpi-hd.mpg.de} 
\\\\
{\normalsize \it$^a$Harish--Chandra Research Institute,}\\
{\normalsize \it Chhatnag Road, Jhunsi, 211019 Allahabad, India }\\ \\ 
{\normalsize \it$^b$Max--Planck--Institut f\"ur Kernphysik,}\\
{\normalsize \it  Postfach 103980, D--69029 Heidelberg, Germany} 
}
\date{}
\maketitle
\thispagestyle{empty}
\vspace{-0.8cm}
\begin{abstract}
\noindent  
Among all neutrino mixing parameters, the
atmospheric neutrino mixing angle 
$\theta_{23}$ introduces 
the strongest variation on the flux ratios of 
ultra high energy neutrinos. 
We investigate the potential  
of these flux ratio measurements at neutrino 
telescopes to constrain 
$\theta_{23}$. We consider astrophysical neutrinos 
originating from pion, muon-damped and 
neutron sources and make a comparative study of their 
sensitivity reach to $\theta_{23}$. It is found that 
neutron sources are most favorable for testing 
deviations from maximal $\theta_{23}$.
Using a $\chi^2$ analysis, 
we show in particular the power of combining (i) different 
flux ratios from the same type of source,  
and also (ii) combining flux ratios from 
different astrophysical sources. 
We include in our analysis 
``impure'' sources, i.e., deviations from the 
usually assumed initial $(1 : 2 : 0)$, $(0 : 1 : 0)$ 
or $(1 : 0 : 0)$ 
flux compositions.

\end{abstract}

\newpage

\section{\label{sec:intro}Introduction}

Dedicated facilities spanning km$^2$ of area 
for detecting ultra high energy neutrino 
coming from astrophysical sources 
are under construction (IceCube \cite{ice,icefirst})
or consideration (KM3Net \cite{km3}). 
Motivated by this, a significant number of papers 
has been devoted in recent years to the phenomenon of neutrino mixing of 
high-energy astrophysical 
neutrinos \cite{LP,fluxes,francesco,GR,Michael,Pasquale,xing2,WW,
xing,KT,old,WR,MO,sterile,CP,kachneu,new,PRW0,PRW1}. 
That neutrinos are massive and therefore mix has been proved 
beyond any doubt by observations of neutrinos coming 
from the Sun \cite{solar}, atmosphere \cite{atm}, 
reactors \cite{kl} and accelerators \cite{k2k,minos}. 
The most recent limits on the mass-squared 
differences and the mixing angles can be found in \cite{limits}. 
The mass splitting associated with solar neutrino oscillations 
is $\Delta m_{21}^2 \simeq 7.6 \times 10^{-5}$ eV$^2$, while 
that associated with atmospheric neutrinos is 
$|\Delta m_{31}^2| \simeq 2.5 \times 10^{-3}$ eV$^2$. 
Because the oscillation lengths corresponding to 
these mass-squared differences are much smaller than 
astrophysical distances, the oscillations get averaged out for the 
astrophysical neutrinos. However, the non-trivial 
flavor mixing in the lepton sector still 
modifies the neutrinos fluxes on their way from the 
source to the detector. This 
opens up the possibility to obtain information 
on the mixing angles and the CP phase of the PMNS matrix. 
This information would be complementary to the already 
impressive existing and expected 
data from current and future experiments devoted purely 
to neutrino oscillations. 

Ultra high energy (UHE) neutrinos are expected to come from 
decay of pions, muons and/or neutrons. 
Therefore, even though the absolute numbers of UHE 
neutrinos are uncertain by a huge amount, the 
relative proportions of the initial 
flux compositions $\Phi_e^0 : \Phi_\mu^0 : \Phi_\tau^0$
are known. Here $\Phi_\alpha^0$ with $\alpha = e, \mu, \tau$ 
is the initial flux of a neutrino with flavor $\alpha$. 
This ratio is $(1 : 2 : 0)$, $(0 : 1 : 0)$ and 
$(1 : 0 : 0)$ for pion, muon-damped and neutron sources, respectively. 
Hence, working with the flux ratios of different flavors 
is considered to be much less model dependent than working with 
absolute fluxes. 
One complication which still arises is that in general 
one expects corrections \cite{lipari,Michael,kachneu} 
to the usually assumed ``pure'' initial 
flux compositions. For instance, instead of 
$(\Phi_e^0 : \Phi_\mu^0 : \Phi_\tau^0) = (1 : 2 : 0)$, 
one might initially 
have $(1 : 2 \, (1 - \zeta) : 0)$, with $\zeta$ being around = 0.1 
\cite{lipari}. Not taking such deviations into account 
can lead to wrong conclusions about the 
neutrino parameters \cite{PRW1}. 

Currently the mixing angle $\theta_{23}$ is one of the least known 
besides $\theta_{13}$, the CP phase and the 
neutrino mass hierarchy. In fact, the quantity $\sin^2\theta_{23}$ 
is uncertain by about $\pm 33\%$ at $3\sigma$. 
Information on this mixing angle and its deviation from maximality 
can be tested in future atmospheric neutrino experiments 
\cite{th23GMS,th23CR}. Using the zenith angle
dependence of the muon events in atmospheric neutrino data, 
one would be able to restrict $\sin^2\theta_{23}$ to 
within $\pm 24\%$ with 1.84 MTy (megaton-year) data in water Cerenkov 
detectors and to within $\pm 30\%$ with 250 kTy data in 
large magnetized iron calorimeters \cite{nu2006talk}. 
Atmospheric neutrino 
data 
could also be used very effectively to give us the 
``octant'' (whether $\theta_{23} < \pi/4$ or $> \pi/4$)
of $\theta_{23}$, if indeed $\theta_{23}$ turns out 
to be non-maximal. 
The sub-GeV electron events in water 
Cerenkov detectors carry information on the 
$\theta_{23}$ octant through the $\Delta m_{21}^2$-driven 
sub-dominant oscillations \cite{th23GMS}. 
The multi-GeV muon events in large magnetized iron 
calorimeters have sensitivity to Earth matter effects which 
in turn depend on $\theta_{23}$ and its octant \cite{th23CR}. 
Long baseline experiments would also give very good sensitivity 
to $\sin^2 \theta_{23}$: the combined 5 year data from 
MINOS, ICARUS, OPERA, T2K and NO$\nu$A is expected to 
constrain $\sin^2\theta_{23}$ to within $\pm 20\%$ 
around its maximal value \cite{huber10}. 

In this letter we will focus on the possibility to constrain  
the atmospheric neutrino mixing angle $\theta_{23}$ from measurements 
of flux ratios at neutrino telescopes. 
After arguing that their dependence 
on this angle is strongest among the neutrino mixing parameters, 
we discuss its possible constraints using a $\chi^2$ analysis.
We quantify that combining different sources leads to stronger  
constraints on $\theta_{23}$. If different flux ratios from 
different neutrino sources are combined (those are 
pion, muon-damped and neutron sources), even better constraints 
are possible. Neutron beam sources turn out to be the most 
interesting ones. 
We include the possibility 
of ``impure sources'' in our analysis and 
investigate their impact for the first time 
in a statistical analysis.

We begin by discussing the flux ratios of UHE at the 
neutrino telescopes in Section 2. In Section 3 we 
introduce our $\chi^2$ function and use it to 
give prospective bounds on $\theta_{23}$ using 
the flux ratios. We end in Section 4 with a 
summary of our results and conclusions.

\section{\label{sec:nutel}Neutrino Mixing and Neutrino Telescopes}
Astrophysical sources will generate fluxes of electron, 
muon and tau neutrinos, 
denoted by $\Phi_e^0$, $\Phi_\mu^0$ and $\Phi_\tau^0$, respectively. 
As a consequence 
of non-trivial neutrino mixing, it is not this initial flux 
composition which arrives at terrestrial detectors.  
In fact, what is measurable is given by 
\be
\left( 
\ba
\Phi_e \\
\Phi_\mu \\
\Phi_\tau 
\ea
\right) 
= 
\left( 
\bad 
P_{ee} & P_{e \mu} & P_{e \tau} \\
P_{\mu e } & P_{\mu\mu} & P_{\mu \tau} \\
P_{\tau e} & P_{\tau \mu} & P_{\tau \tau} 
\ea
\right) 
\left( 
\ba
\Phi_e^0 \\
\Phi_\mu^0 \\
\Phi_\tau^0 
\ea
\right)~,
\ee
where the neutrino mixing probability is 
\be 
P_{\alpha \beta} = P_{\beta \alpha} = \sum\limits_i |U_{\alpha i}|^2\, 
|U_{\beta i}|^2~ 
\ee
and $U$ is the lepton mixing matrix.
The flavor mixing matrix $P$ comprises of the 
individual $P_{\alpha \beta}$ elements. 
The current best-fit values as well as the 
allowed $ 1\sigma$ and $3\sigma$ ranges of the 
oscillation parameters are \cite{limits}: 
\begin{eqnarray} \label{eq:data}
\sin^2 \theta_{12}
 &=& 0.32^{+0.02\,, \,0.08}_{-0.02\,, \,0.06} ~, \nonumber \\
\sin^2\theta_{23} &=& 0.45^{+0.09\,, \,0.19}_{-0.06\,, \,0.13} ~,\\
\sin^2\theta_{13}  &<& 0.019~(0.050)~.\nonumber
\end{eqnarray}
These mixing angles can be related to elements of the PMNS 
mixing matrix via 
\bea \label{eq:Upara}
U = 
\left( \bad 
c_{12} \, c_{13} & s_{12}   \, c_{13} & s_{13}  \, e^{-i \delta} \\[0.2cm] 
-s_{12}   \, c_{23} - c_{12}   \, s_{23}   \, s_{13}   \, e^{i \delta}  
& c_{12}   \,  c_{23} - s_{12}  \,   s_{23}  \,   s_{13}  \,  e^{i \delta}
& s_{23}   \,  c_{13}  \\[0.2cm] 
s_{12}  \,   s_{23} - c_{12}  \,   c_{23}  \,   s_{13}  \, e^{i \delta}& 
- c_{12}  \,   s_{23} - s_{12}  \,   c_{23}   \,  s_{13} \,    e^{i \delta}
& c_{23}   \,  c_{13}  \\ 
               \ea   \right)  ~, 
\eea
where $c_{ij} = \cos\theta_{ij}$, 
$s_{ij} = \sin\theta_{ij}$. 
The CP phase $\delta$ is unknown. Because the mass-squared differences 
drop out of the mixing probabilities, and solar neutrino mixing is 
neither maximal, zero nor $\pi/2$, no transition probability 
$P_{\alpha \beta}$ with $\alpha \neq \beta$ is zero 
and no survival probability $P_{\alpha \alpha}$ is one. 
Consequently, high energy astrophysical neutrinos will always mix. 
To be precise, at $1\sigma$ and $3 \sigma$ the entries of the 
flavor conversion matrix $P_{\alpha \beta}$ are 
\be \label{eq:ranges}
P  = 
\left\{
\baz 
\left( 
\bad 
0.53 \div 0.58 & 0.18 \div 0.30 & 0.16 \div 0.27 \\
\cdot & 0.34 \div 0.44 & 0.35 \div 0.40 \\
\cdot & \cdot & 0.35 \div 0.47
\ea
\right) & (\mbox{at } 1\sigma)~,\\[0.3cm]
\left( 
\bad 
0.47 \div 0.62 & 0.12 \div 0.35 & 0.11 \div 0.34 \\
\cdot & 0.33 \div 0.51 & 0.30 \div 0.40 \\
\cdot & \cdot & 0.33 \div 0.53
\ea
\right) & (\mbox{at }3\sigma)~.
\ea
\right. 
\ee
As is obvious from Eq.~(\ref{eq:data}), a good zeroth order description 
of neutrino mixing is tri-bimaximal mixing \cite{tbm},  
\be \label{eq:tbm}
U \simeq U_{\rm TBM} = 
\left( 
\bad
\sqrt{\frac{2}{3}}  &  \frac{1}{\sqrt{3}}  & 0 \\[0.cm]
-\frac{1}{\sqrt{6}} 
& \frac{1}{\sqrt{3}}  & -\frac{1}{\sqrt{2}}\\[0.cm]
-\frac{1}{\sqrt{6}} & \frac{1}{\sqrt{3}}  & \frac{1}{\sqrt{2}}
\ea 
\right)~.
\ee
Therefore, it proves in particular useful to expand 
in terms of\footnote{An expansion 
up to second order around zero $|U_{e3}|$, 
maximal $\theta_{23}$ 
and $\sin^2 \theta_{12} = \frac 13$ 
can be found in Refs.~\cite{PRW0,PRW1}.}
\be 
|U_{e3}|\mbox{ and } 
\epsilon \equiv \frac{\pi}{4} - \theta_{23} = 
\frac 12 - \sin^2 \theta_{23} + {\cal O}(\epsilon^3)~.
\ee
For simplicity, 
we fix from now on for analytical considerations 
$\sin^2 \theta_{12}$ to $\frac 13$. This is a valid approximation  
for obtaining simple analytical expressions, but we stress that the  
numerical results to be presented in this paper are 
obtained with the exact expressions. The result for the 
flavor mixing matrix is 
\bea \label{eq:res1a}
P 
\simeq   
\left( 
\bad
\frac 59 &   \frac 29 &  \frac 29  \\
 \cdot & \frac{7}{18} &  \frac{7}{18}  \\
 \cdot & \cdot  
& \frac{7}{18}  
\ea
\right)
+ \Delta 
\left( 
\bad
0 & 1 & -1 \\ 
\cdot & -1 & 0 \\ 
\cdot & \cdot & 1
\ea
\right) 
+ 
\frac 12 \, \overline{\Delta}^2
\left( 
\bad
0 & 0 & 0 \\ 
\cdot  & 1 & -1 \\ 
\cdot  & \cdot  & 1
\ea
\right)~,
\ea
\ee
where the universal first \cite{xing,WR} and second \cite{PRW0,PRW1} 
order correction terms are 
\bea \D  \label{eq:Del}
\Delta = \frac 19 
\left( 
\sqrt{2} \, \cos \delta \, |U_{e3}| + 4 \, \epsilon
\right)~,\\[0.2cm] \D 
\overline{\Delta}^2 = \frac 49 
\left( 
2 \cos^2 \delta \, |U_{e3}|^2 + 7 \, \epsilon^2 - \sqrt{2} 
\, \cos \delta \, |U_{e3}| \, \epsilon
\right)~.
\eea
We have omitted here small and usually negligible terms in the expansion 
which only contain $|U_{e3}|^2$, see Refs.~\cite{PRW0,PRW1}. 
Note that $\overline{\Delta}^2$ can be written as 
$\frac 19 \, (2\sqrt{2} \, \cos \delta \, |U_{e3}| - \epsilon)^2 + 
3 \, \epsilon^2$ and therefore is positive semi-definite. In 
addition, $\overline{\Delta}^2$ turns out to be often larger than 
the first order term $\Delta$ \cite{PRW0,PRW1}. To be quantitative, 
\be \label{eq:range2}
\baf   
\mbox{at } 1\sigma:& -0.043 \le \Delta \le 0.069 & \mbox{, } 
& \overline{\Delta}^2 \le 0.061 ~,\\[0.2cm]
\mbox{at } 3\sigma:& -0.104 \le \Delta \le 0.117 & \mbox{, } & 
\overline{\Delta}^2 \le 0.179~.
\ea
\ee
The dependence of $\Delta$ and $\overline{\Delta}^2$ 
on $\theta_{12}$ is very weak \cite{PRW0,PRW1}. 
From the expressions for $\Delta$ and $\overline{\Delta}^2$ it is clear 
that their dependence on the atmospheric 
neutrino mixing angle $\theta_{23}$ is stronger than 
on $\cos \delta \, |U_{e3}|$. It is the large prefactor in front of 
$\epsilon$ and $\epsilon^2$ which is the reason for this behavior. 
In addition, these prefactors are larger than the ones related to 
$|U_{e3}|$. Hence, the dependence on $|U_{e3}|$ is weaker and 
smeared by the additional dependence on $\cos \delta$. 
This is why we are interested here in effects 
of deviations from maximal atmospheric neutrino 
mixing.\\ 

The neutrino sources we assume are pion, muon-damped \cite{010} 
and neutron \cite{100} sources, with initial flux compositions of 
\be \label{eq:sources}
(\Phi_e^0 : \Phi_\mu^0 : \Phi_\tau^0) = 
\left\{
\baz 
(1 : 2 \, (1 - \zeta) : 0) &  \mbox{pion} ~,\\
(\eta : 1 : 0) &  \mbox{muon-damped} ~,\\
(1 : \eta : 0) &  \mbox{neutron}~.
\ea
\right.
\ee
By introducing small $\zeta, \eta \ll 1$ \cite{PRW1}, we have 
allowed here for 
impurities in the initial compositions, which can 
be expected on general grounds \cite{lipari,Michael,kachneu}. \\

\begin{figure}[ht]
\begin{tabular}{cc}
\includegraphics[width=8cm,height=7cm]{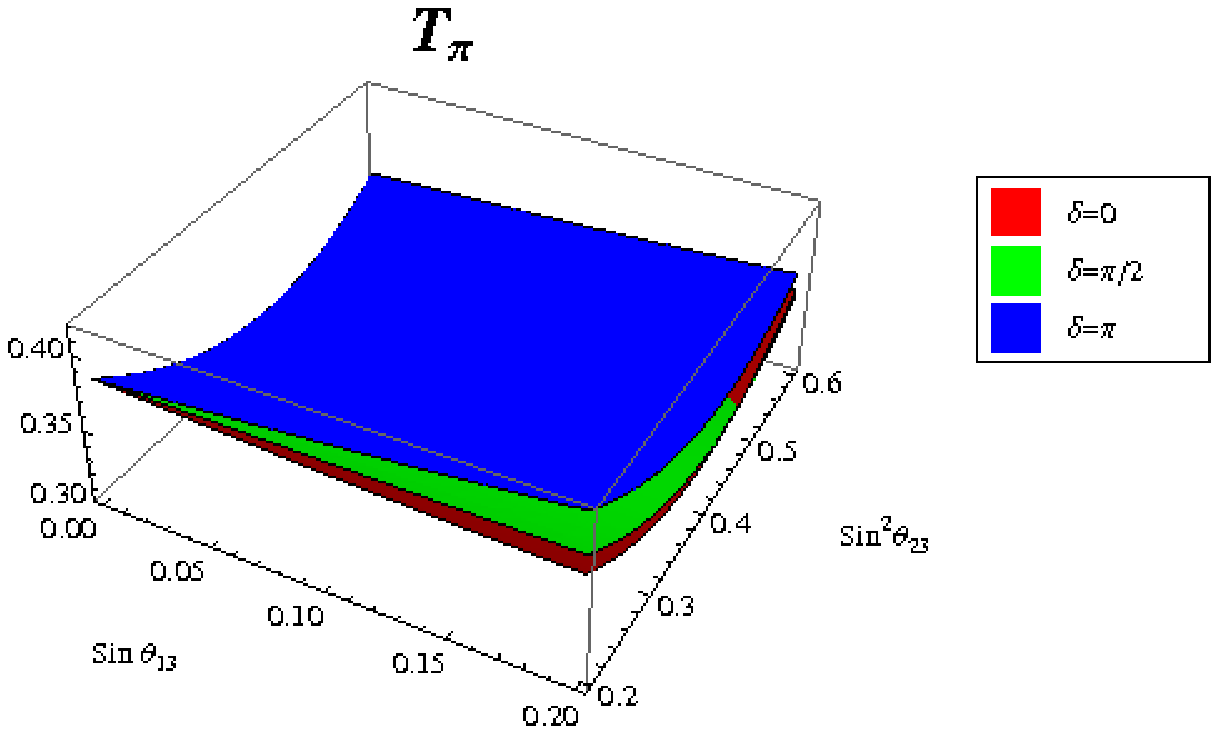} & 
\includegraphics[width=8cm,height=7cm]{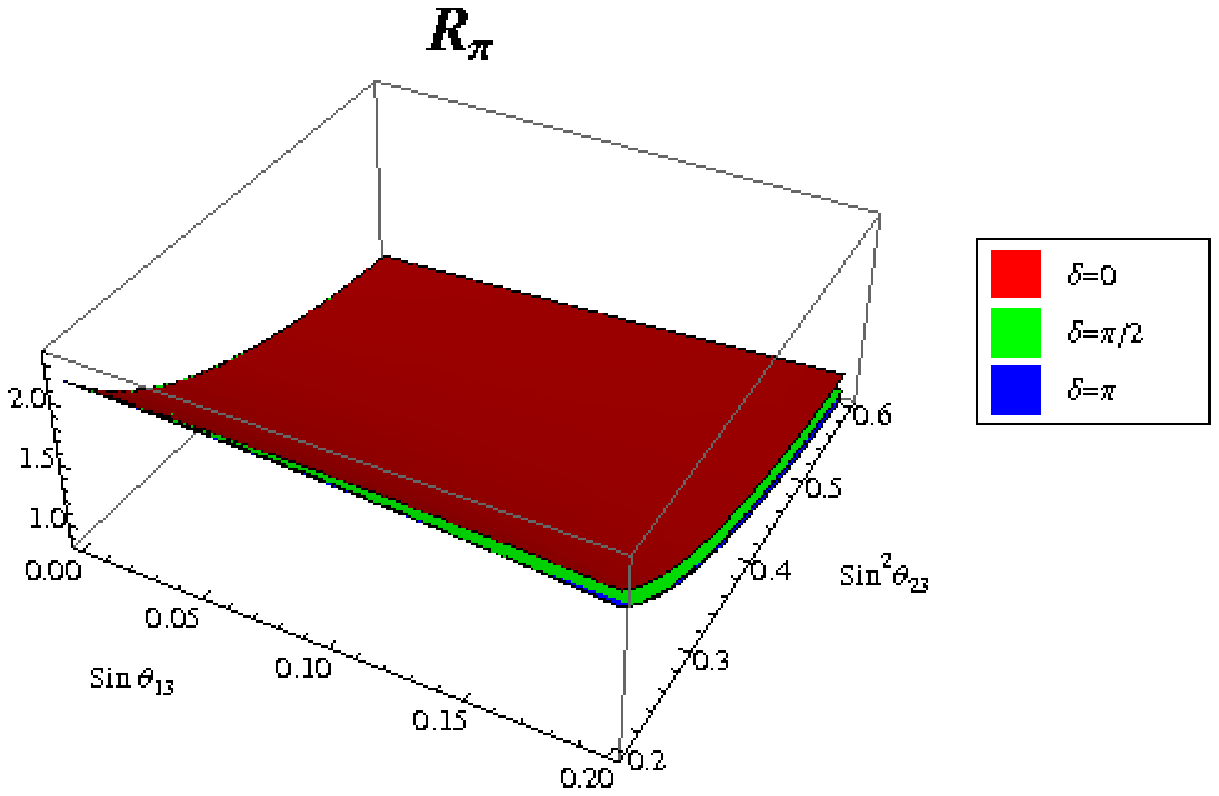} \\ 
\includegraphics[width=8cm,height=7cm]{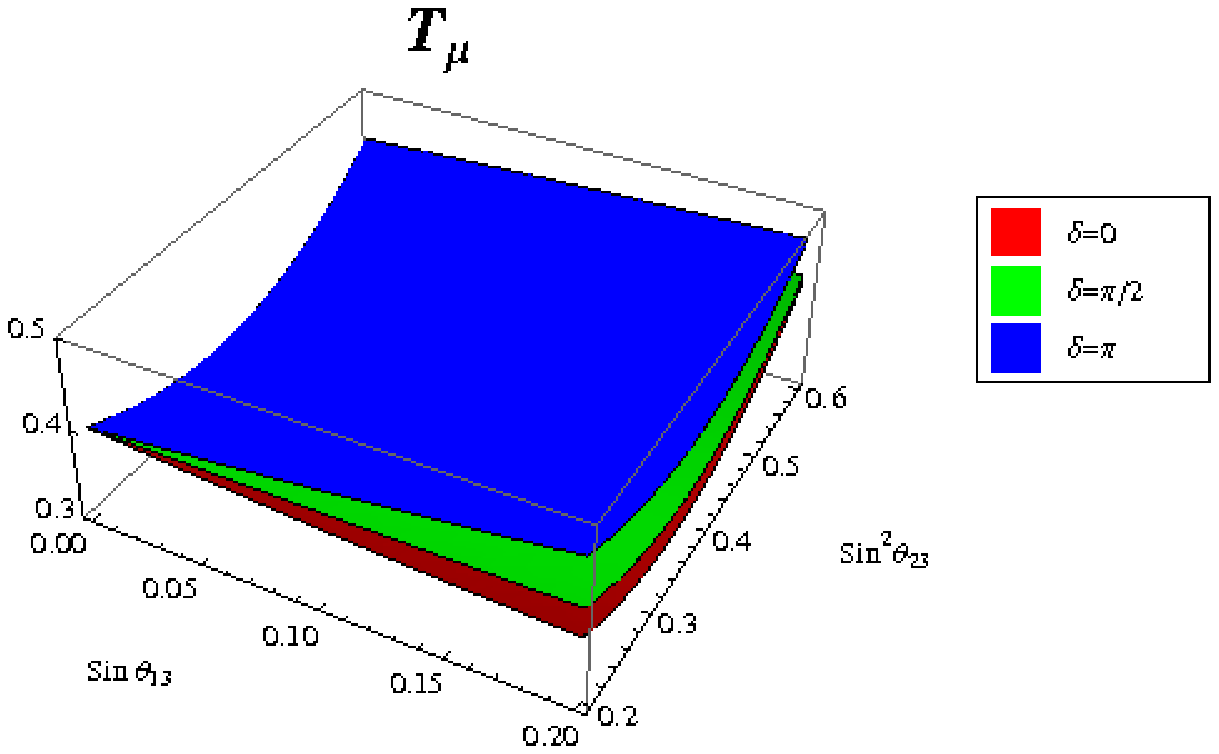} & 
\includegraphics[width=8cm,height=7cm]{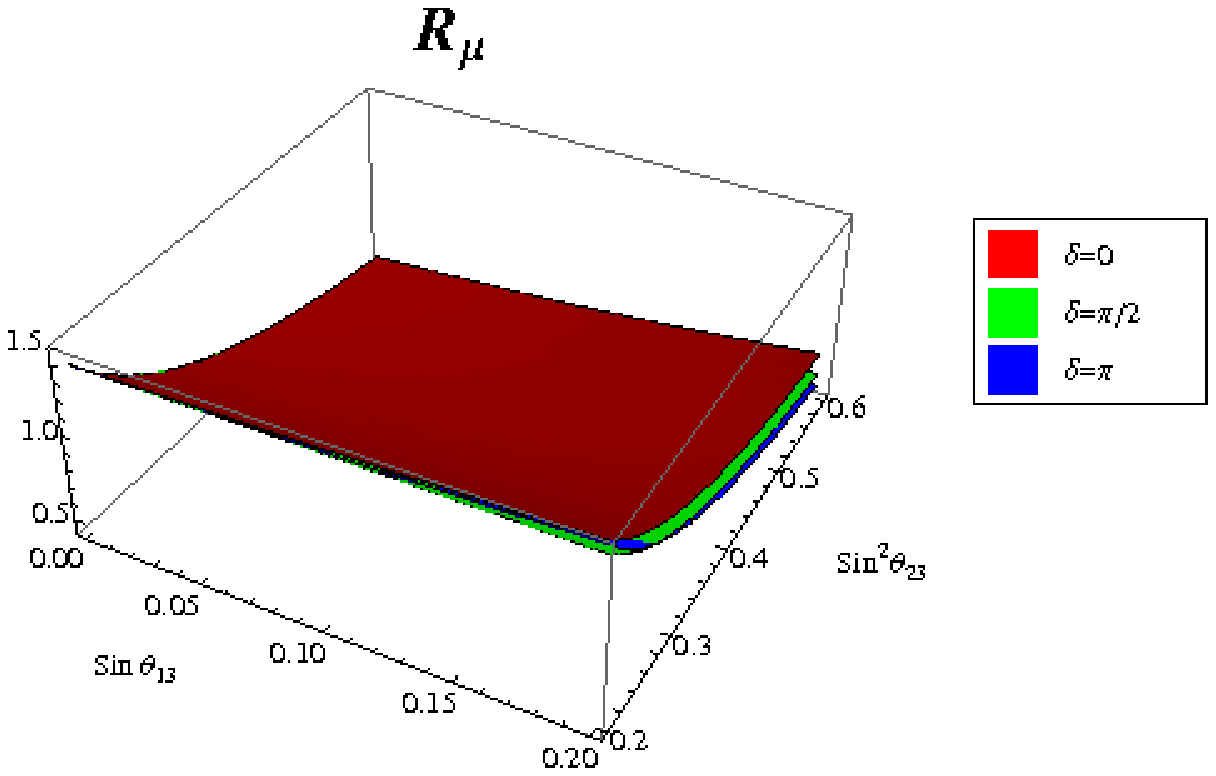} \\
\includegraphics[width=8cm,height=7cm]{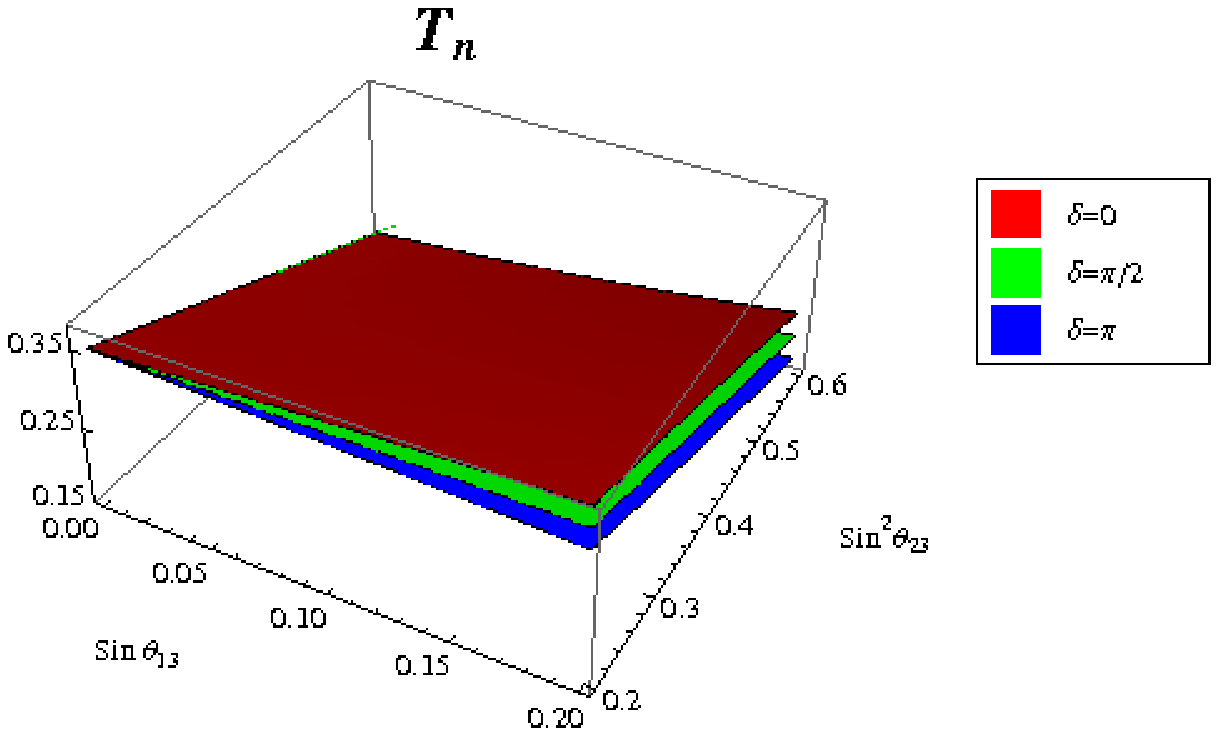} & 
\includegraphics[width=8cm,height=7cm]{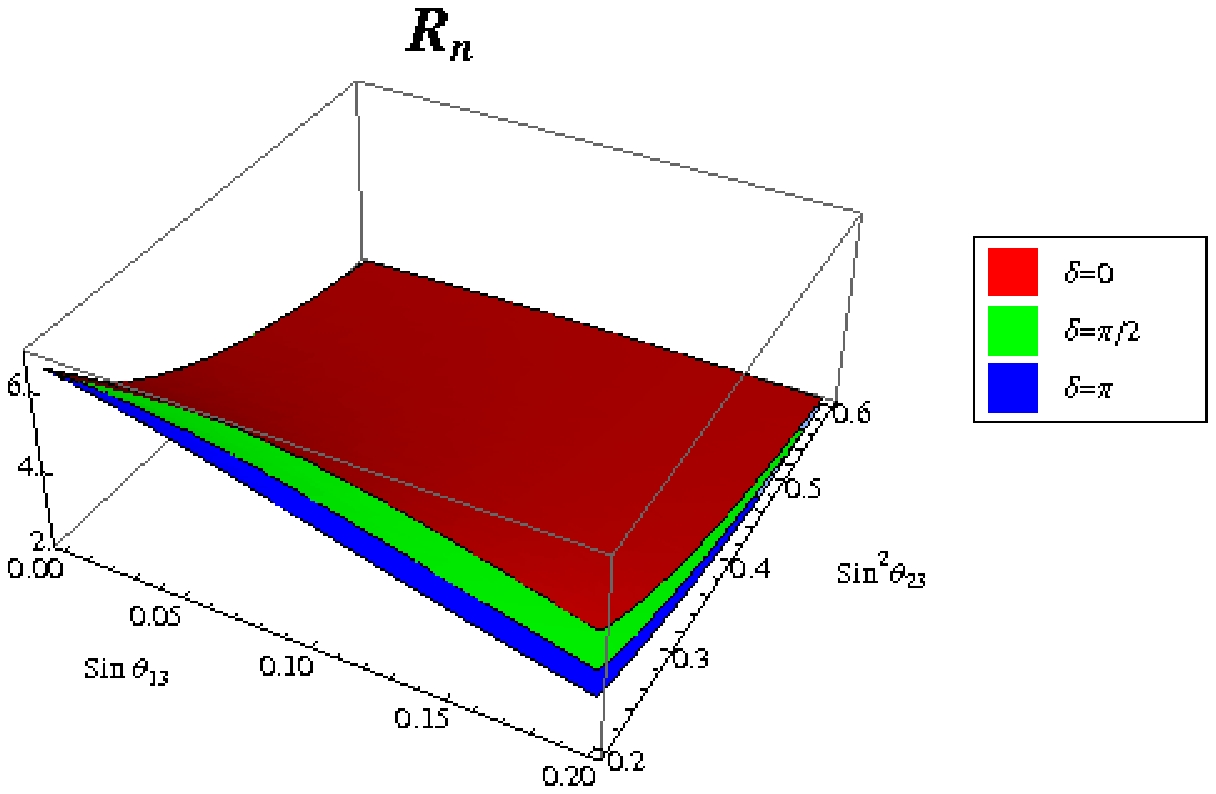}
\end{tabular}
\caption{\label{fig:3d}The ratios $T = 
\Phi_\mu/\Phi_{\rm tot}$ (left panel) and $R = \Phi_e/\Phi_{\tau}$ 
(right panel) 
for the three sources under discussion as a function of 
$\sin \theta_{13}$ and $\sin^2 \theta_{23}$. 
The mixing angle $\theta_{12}$ 
has been fixed to its best-fit value, i.e., 
$\sin^{2}\theta_{12} = 0.32$.}
\end{figure}

\begin{figure}[ht]
\begin{tabular}{cc}
\includegraphics[width=8cm,height=6cm]{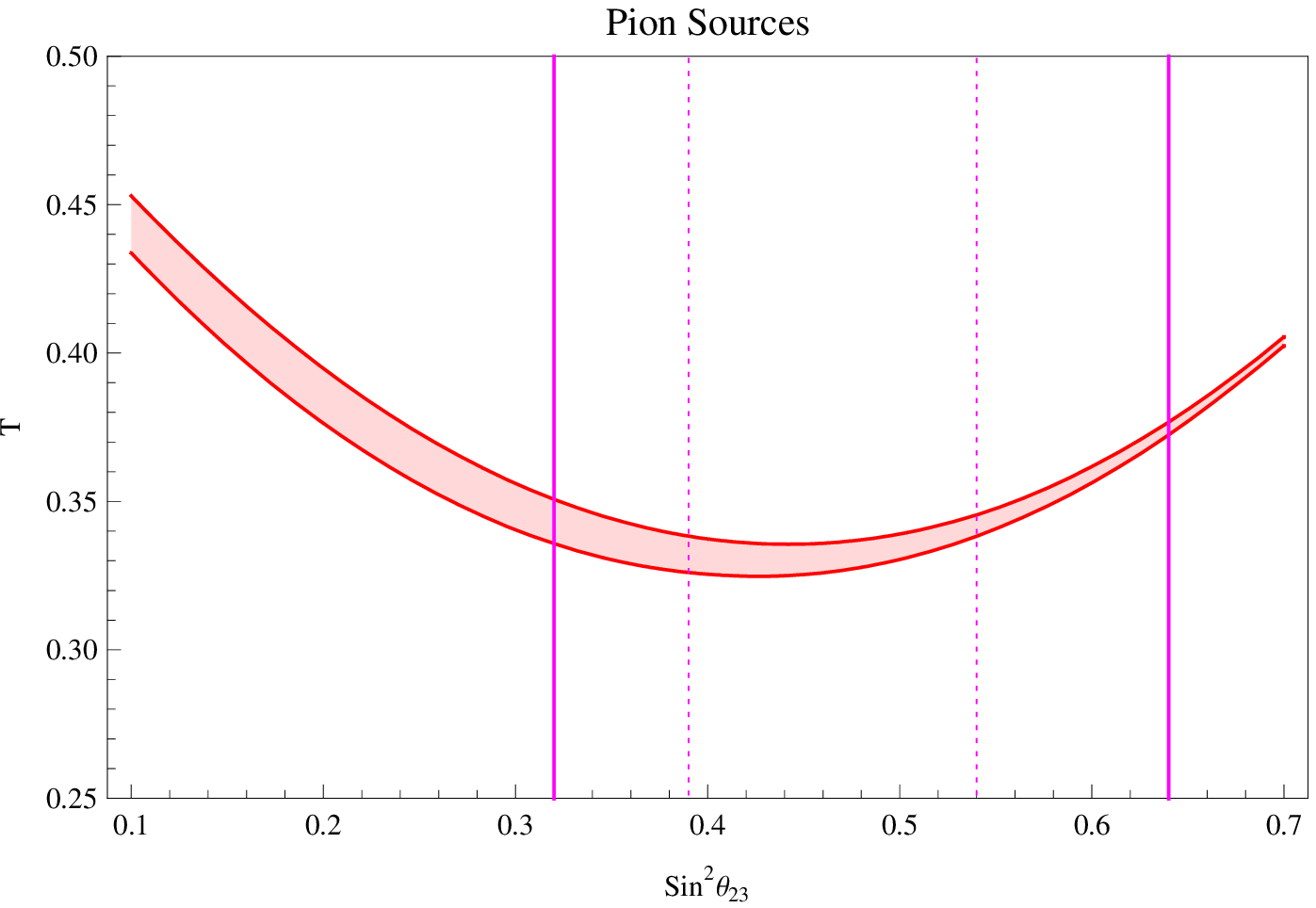} & 
\includegraphics[width=8cm,height=6cm]{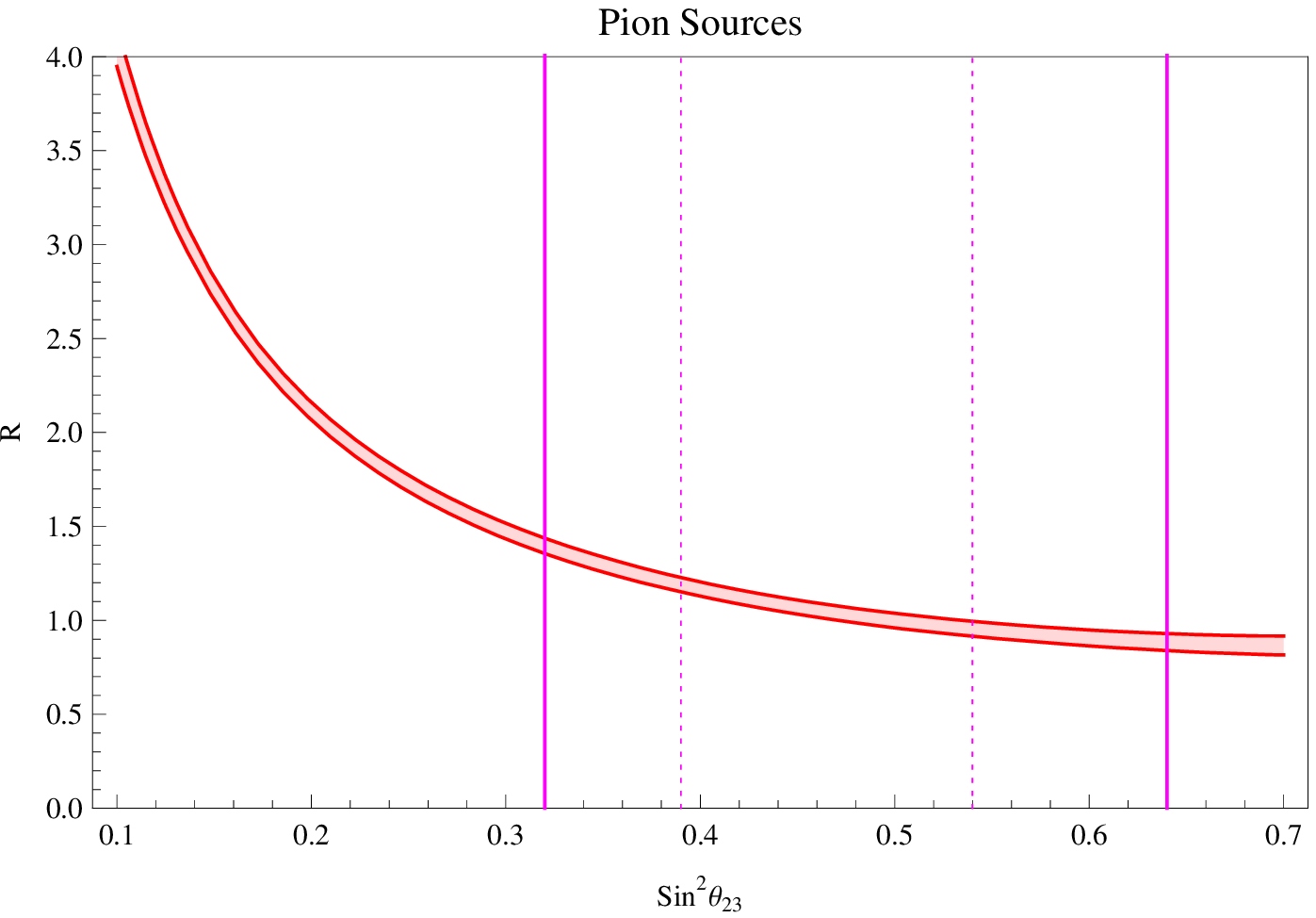} \\
\includegraphics[width=8cm,height=6cm]{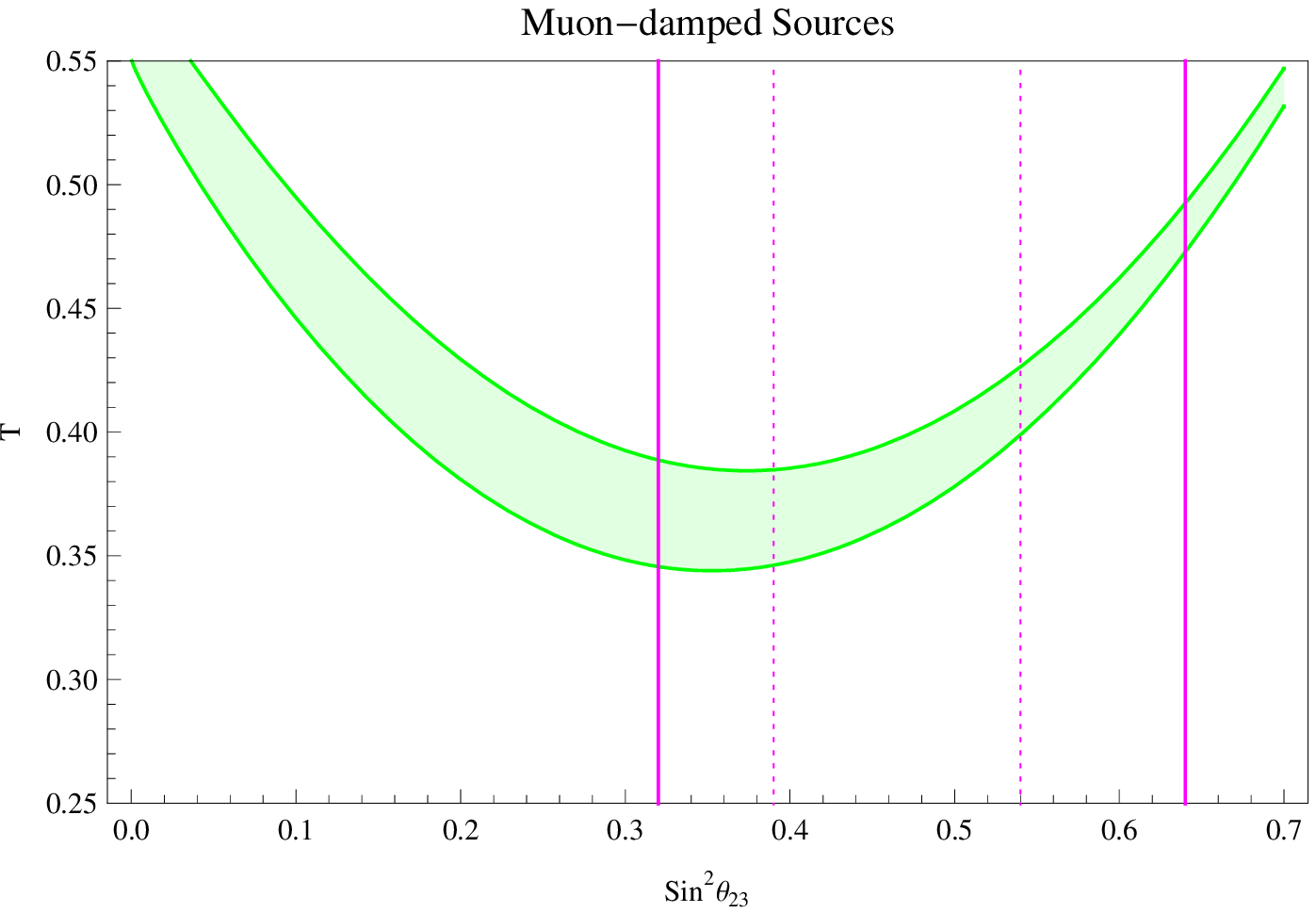} & 
\includegraphics[width=8cm,height=6cm]{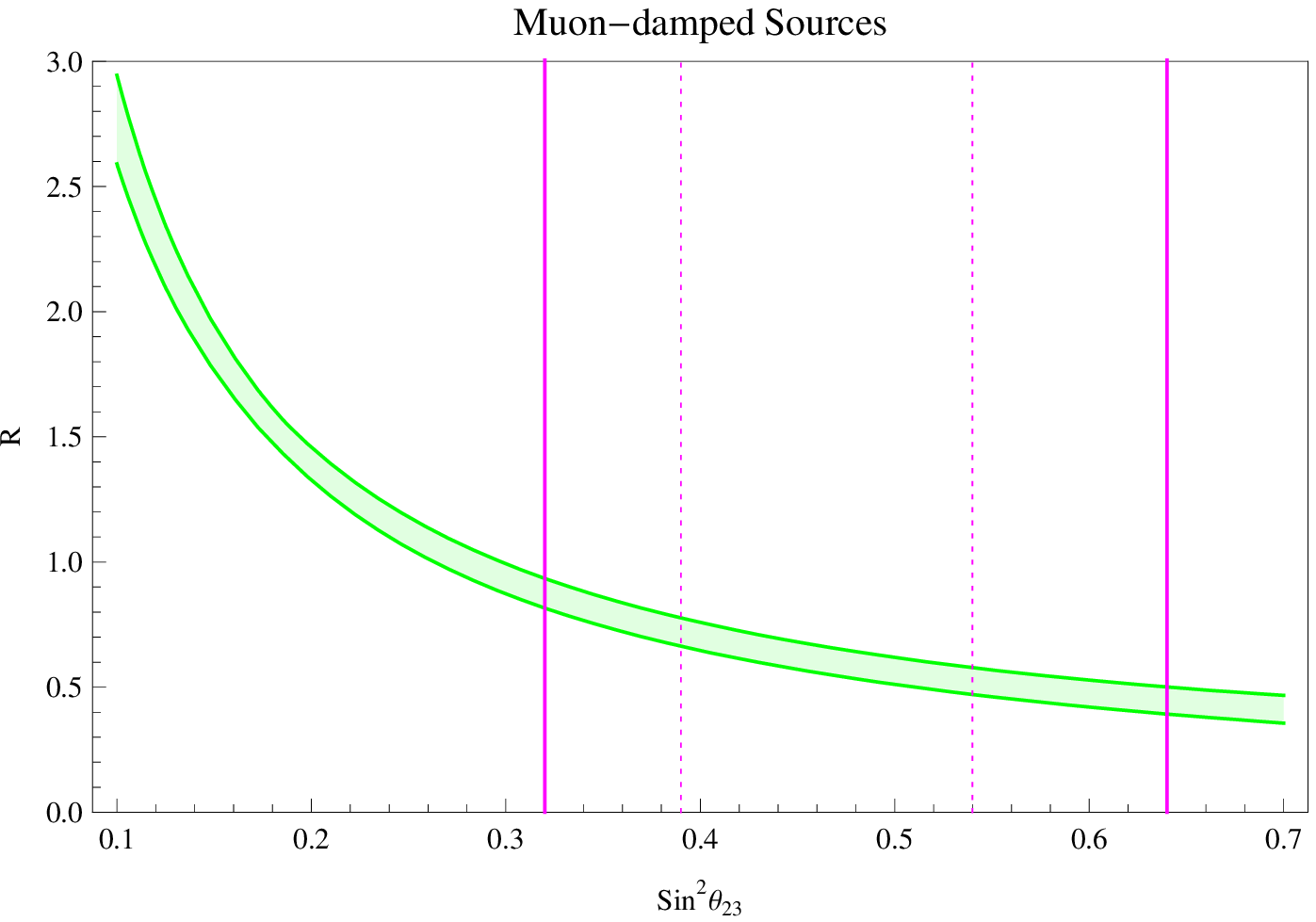} \\
\includegraphics[width=8cm,height=6cm]{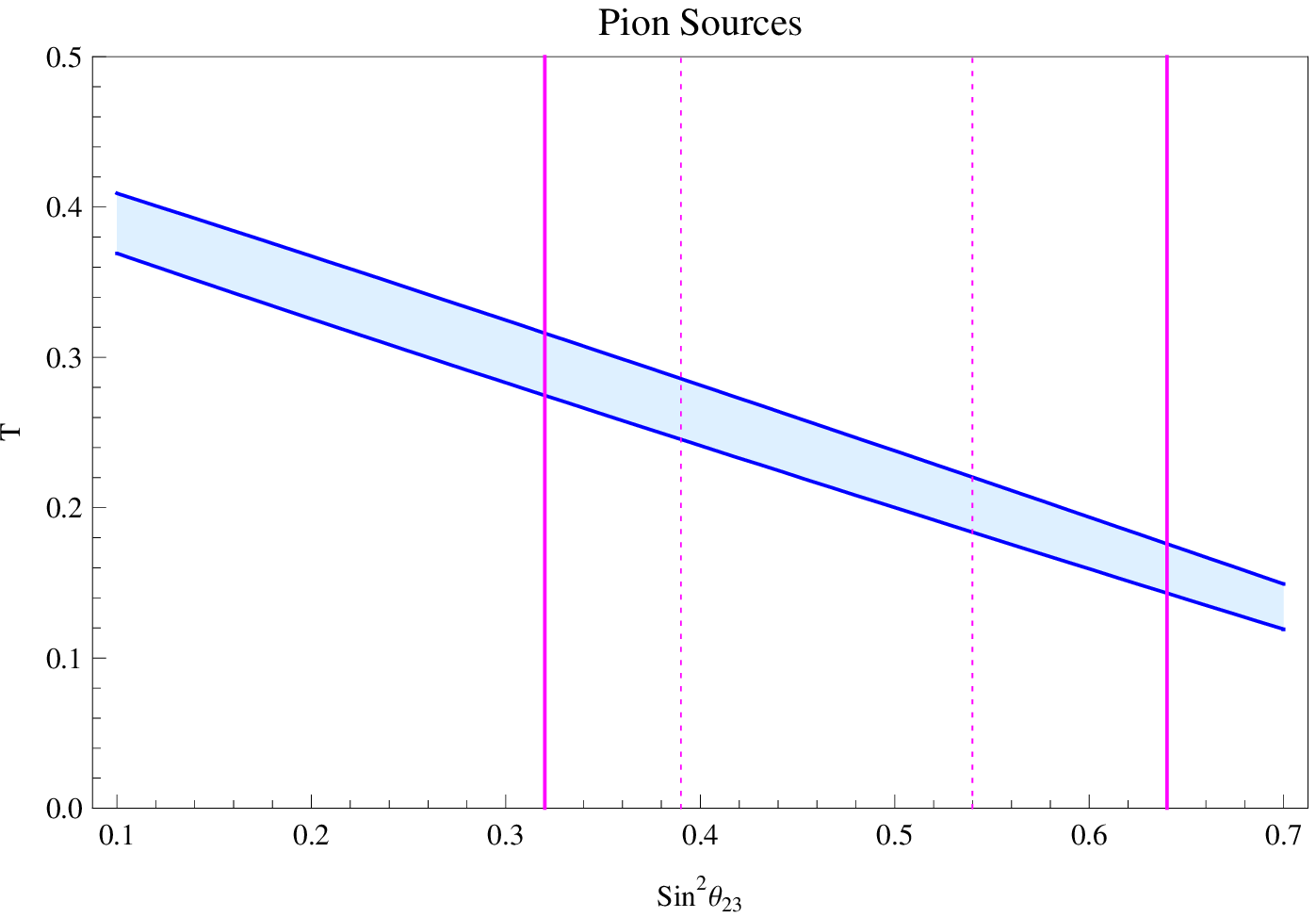} & 
\includegraphics[width=8cm,height=6cm]{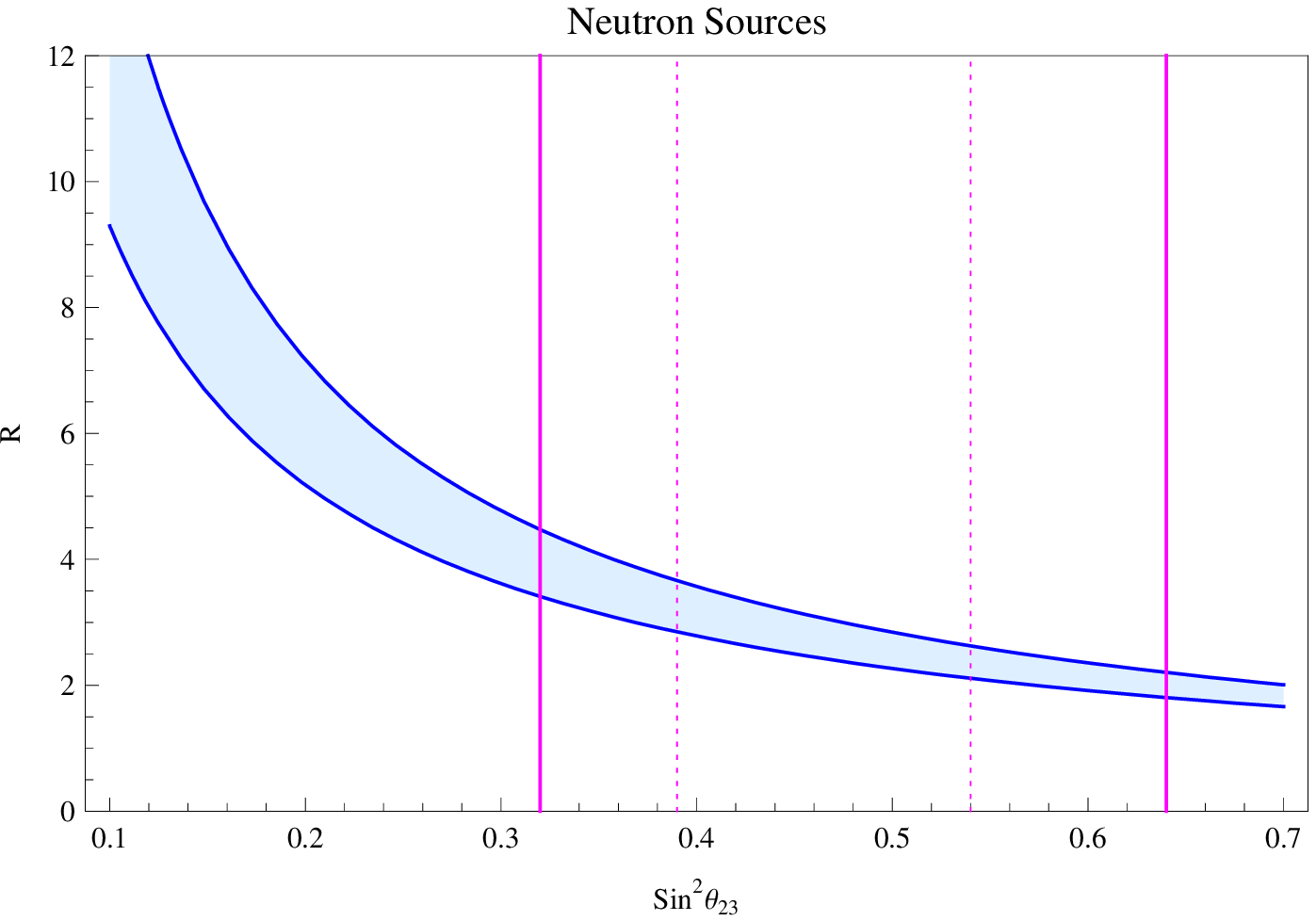}
\end{tabular}
\caption{\label{fig:2d}The ratios $T = 
\Phi_\mu/\Phi_{\rm tot}$ (left panel) and $R = \Phi_e/\Phi_{\tau}$ 
(right panel) for the three sources as a function of 
$\sin^2 \theta_{23}$ when the ranges of 
the mixing parameters are $\sin^{2}\theta_{12} = 0.32 \pm 0.02$ and 
$\sin^{2}\theta_{13} \le 0.005$. 
The current $1\sigma$ and 
$3\sigma$ ranges of $\sin^2 \theta_{23}$ are also indicated.}
\end{figure}

Turning to the observable flavor ratios \cite{ratio1,ratio2}, 
the most frequently 
considered is the ratio of 
muon neutrinos to all other flavors: 
\be \label{eq:defT}
T = \frac{\Phi_\mu}{\Phi_e + \Phi_\mu + \Phi_\tau} 
= \frac{\Phi_\mu}{\Phi_{\rm tot}}~.
\ee
Using again $\sin^2 \theta_{12} = \frac 13$ and 
expanding in terms of the small parameters $|U_{e3}|$, 
$\epsilon = \frac{\pi}{4} - \theta_{23}$ as well as $\zeta$ or $\eta$ 
one finds \cite{PRW1}
\be \label{eq:T}
T \simeq  
\left\{ 
\baz 
\frac 13 \left( 1 - \Delta + \overline{\Delta}^2 
- \frac 19 \, \zeta \right)\,,
& \mbox{pion }  (1 : 2 \, (1 - \zeta) : 0)~,  \\[0.2cm]
\frac{7}{18} - \Delta + \frac 12 \, \overline{\Delta}^2 - 
\frac 16 \, \eta\,, & \mbox{muon-damped } (\eta : 1 : 0)~, \\[0.2cm]
\frac 29 + \Delta + \frac 16 \, \eta\,, & 
\mbox{neutron beam }  (1 : \eta : 0)~.
\ea 
\right.
\ee
The second order correction $\overline{\Delta}^2$ appears only in 
$P_{\mu\mu}$, $P_{\mu\tau}$ and $P_{\tau\tau}$, and therefore 
does not affect $T$ for neutron sources. 
Hence, the dependence on $\sin^2 \theta_{23}$ 
can be described to an excellent approximation as quadratic for 
pion and muon-damped sources, but linear for neutron sources. 
We show in the left panel 
of Fig.~\ref{fig:3d} the ratio $T$ 
as a function of $\sin^2 \theta_{23}$ and $\sin \theta_{13}$ 
for the three pure sources (i.e., $\zeta = \eta = 0$). 
Plots are shown for fixed $\sin^2\theta_{12}=0.32$ and 
three values of $\delta$. 
The stronger 
dependence on $\sin^2 \theta_{23}$ is clearly seen. 
Fig.~\ref{fig:2d} 
focusses on the dependence of the flux ratios 
on $\sin^2 \theta_{23}$. 
We have taken the range 
$\sin^2\theta_{12}=0.32 \pm 0.02$, $\sin^2\theta_{13}<0.005$
and $0 \le \delta < 2\pi$. 
We have also marked the current 
$1 \sigma$ and  $3 \sigma$ allowed ranges of $\sin^2\theta_{23}$ 
for comparison. 
We reiterate that these plots are for $\zeta = \eta = 0$.
Note that because of larger prefactors in front of 
$\zeta$ and $\eta$ in Eq.~(\ref{eq:T}), 
impurities are expected to have stronger 
impact for muon-damped and neutron sources. 
Finally, we stress that the correction 
factors 
$\Delta$ and $\overline{\Delta}^2$, which can both be 
${\cal O}(0.1)$, are added to a number around 1 for pion sources, 
but to a number around 0.4 for muon-damped and 0.2 for 
neutron sources. Hence, 
neutron sources are expected to be the most useful to constrain 
$\theta_{23}$.

\noindent
Another ratio which can be considered is the ratio of electron to 
tau neutrinos, 
\be \label{eq:defR}
R = \frac{\Phi_e}{\Phi_\tau}~.
\ee
For the three neutrino sources one finds \cite{PRW1}
\be \label{eq:R}
R = \frac{\Phi_e}{\Phi_\tau}\simeq \left\{ 
\baz \nonumber 
1 + 3 \, \Delta \, (1 + \Delta)+ \overline{\Delta}^2 + 
\frac 13 \, \zeta \,,
& \mbox{pion}~ (1 : 2 \, (1 - \zeta) : 0)~, \\[0.2cm]
\frac{4}{7} \left( 
1 + \frac 92 \, \Delta + \frac{9}{7} \,  \overline{\Delta}^2 
+ \frac{27}{14} \,  \eta 
\right)\,,
& \mbox{muon-damped}~ (\eta : 1 : 0)~, \\[0.2cm]
\frac 52 \left( 
1 + \frac 92 \, \Delta - \frac{27}{20} \, \eta 
\right) \,,
& \mbox{neutron beam}~  (1 : \eta : 0)~.
\ea
\right.
\ee
Comparing these expressions with the ones for $T$ in Eq.~(\ref{eq:T}), 
we note that in case of pion and muon-damped sources 
the first and second order correction terms $\Delta$ 
and $\overline{\Delta}^2$ are added in $R$, whereas they are 
subtracted in $T$. Recalling that $\epsilon$ and $\epsilon^2$ 
appear with equal sign in $\Delta$ 
and $\overline{\Delta}^2$, we expect stronger dependence 
on $\theta_{23}$ in $R$ than in $T$. This is illustrated in the 
right panels of 
Figs.~\ref{fig:3d} and \ref{fig:2d}, where we display $R$ 
as a function of $\sin \theta_{13}$ and $\sin^2 \theta_{23}$.  
Again, the dependence on $\sin^2 \theta_{23}$ is basically quadratic for 
pion and muon-damped sources, whereas it is linear for neutron beams. 
The effect of impurity will be in general larger for $R$, because of 
the larger prefactors in front of $\zeta$ and $\eta$. Note also that  
the parameters $\zeta$ and $\eta$ appear in $R$ 
with opposite sign compared to $T$.

\section{\label{sec:ana}Statistical Analysis}

To statistically investigate the prospects of constraining 
$\theta_{23}$ with neutrino telescopes, 
we turn to a $\chi^2$ analysis. First we discuss the case in which only 
$T$ is measured. Let us define the 
$\chi^{2}$ function to be minimized as:
\begin{equation} \label{eq:chi21}
\chi^{2} = \left(\frac{T_{\rm th}-T_{\rm exp}}
{\sigma_{T_{\rm exp}}}\right)^{2} 
+ 
\sum_{ij=12,13} 
\left(\frac{s^{2}_{ij}-(s^{2}_{ij})_{\text{best-fit}}}
{\sigma_{s^{2}_{ij}}}\right)^{2} 
~,
\end{equation}
where $T_{\rm th}$ and $T_{\rm exp}$ are the theoretically 
predicted and experimentally measured $T$ ratios, respectively, and  
${\sigma_{T_{\rm exp}}}$ is the $1\sigma$ uncertainty 
of the measured value $T_{\rm exp}$ in the neutrino telescope. 

We take into account the possible uncertainty of the initial 
flux composition by constructing different $\chi^2$ functions 
for different initial 
flavor compositions and then define the allowed region of 
$\sin^{2}\theta_{23}$ as the maximum range 
obtained by combining the different functions.
The errors on $T$ for the different sources are 
expected to lie in the range 15-25\% after a decade of 
of running of the experiment. It is expected that the 
error on pion sources will be less than those on muon-damped and 
neutron sources. Therefore, 
we present most of our results for fixed assumed 
errors on the flux ratio $T$ as  
$10\%$ for pion, $20\%$ for muon-damped 
and $15\%$ for neutron sources. 
In what regards $R$, the assumed errors are 
$15\%$ for pion, $25\%$ for muon-damped and $20\%$ for neutron sources. 
In general one expects larger errors for $R$ than on $T$:  
muon neutrinos can be identified 
at neutrino telescopes for energies greater than 100 GeV through 
characteristic muon tracks. Electron and tau neutrinos,  
can be identified through showers and effects like ``double bang'' 
events, respectively, only for higher energies, above $10^{6}$ GeV. 
Recall that muon-damping is expected to happen in a generic 
pion source for a particular and limited energy range, in which 
the muon is absorbed before decaying \cite{010}.  Hence, 
less statistics for muon-damped sources with respect to pion sources 
is expected.
Concerning neutron sources, we have chosen an error 
greater than the one for pion sources, because they are expected 
to be characterized in an energy range covered by  
atmospheric neutrino background \cite{100}. 
Therefore the systematic error in 
this case might be greater than the one for pion sources. 
Hence, we have chosen a hierarchy among the errors for the different 
types of sources. 
However, the values for the errors taken above are 
just one choice for the errors, which could be 
different from what has been chosen above. To take into this fact and 
to show the impact of the errors on our analysis, 
we will also show the results as a function of the 
error on $T$. 

We generate the prospective data $T_{\rm exp}$ for two sets 
of mixing angles which we assume as ``true'' and minimize 
the $\chi^2$ function to obtain bounds on the measured 
$\theta_{23}$ at neutrino telescopes. 
In the fit we allow $\theta_{12}$, 
$\theta_{13}$ and $\delta$ to take any value in their physically 
allowed ranges. 
The second term in Eq. (\ref{eq:chi21}) 
takes into account the ``priors'' on the 
mixing angles $\theta_{12}$ and $\theta_{13}$, on which we 
expect better constraints by the time we get the data on 
UHE neutrinos at neutrino telescopes. For the $1\sigma$ 
uncertainty on $\theta_{12}$ we use the range given 
in Eq.~(\ref{eq:data}), while for $\theta_{13}$ we use 
the following upper bound:
\be
\sin^{2} 2\theta_{13} < 0.03 \text{~at 90\% C.L.}~,
\ee
which corresponds to the absence of a signal in the 
Double CHOOZ experiment after three years of operation with 
both detectors \cite{DC}. We have considered in our analysis 
this specific case and we have studied the consequences that 
can be inferred on the atmospheric mixing angle. This situation 
has be considered as an optimistic scenario. Indeed, larger values of 
$\theta_{13}$ than the ones we have chosen 
would worsen the sensitivity on $\theta_{23}$.

We generate $T_{\rm exp}$ and show results for two sets of 
true values for the mixing parameters:
\begin{enumerate}
\item 
$\sin^{2} \theta_{12}=1/3~,~~ \sin^{2} \theta_{13}=0~~
\text{and}~~\sin^{2} \theta_{23}=1/2~$ (Scenario TBM)~,
\item 
$\sin^{2} \theta_{12}=0.32~,~~ \sin^{2} \theta_{13}=0~~
\text{and}~~\sin^{2} \theta_{23}=0.6~$ (Scenario 2)~.
\end{enumerate}
Scenario TBM gives $T_{\rm exp}$ (and $R_{\rm exp}$)
corresponding to their ``tri-bimaximal values'', i.e., 
the zeroth order terms from Eqs.~(\ref{eq:T}) and (\ref{eq:R}). 
Scenario 2 gives $T_{\rm exp}=$ 
0.36, 0.45, and 0.17 respectively for the pion, muon-damped 
and neutron 
sources. The corresponding values of $R_{\rm exp}$ are 
given as 0.90, 0.45, and 2.16 respectively. Scenario 
TBM corresponds to the case where the true value 
of $\theta_{23}$ is maximal, while Scenario 2 
exemplifies a situation where the true value of $\theta_{23}$ 
is non-maximal and $> \pi/4$. We checked that the 
results for $\theta_{23} < \pi/4$ share the same features as  
the ones for $\theta_{23} > \pi/4$.

\begin{figure}[ht]
\begin{tabular}{cc}
\includegraphics[width=8cm,height=6cm]{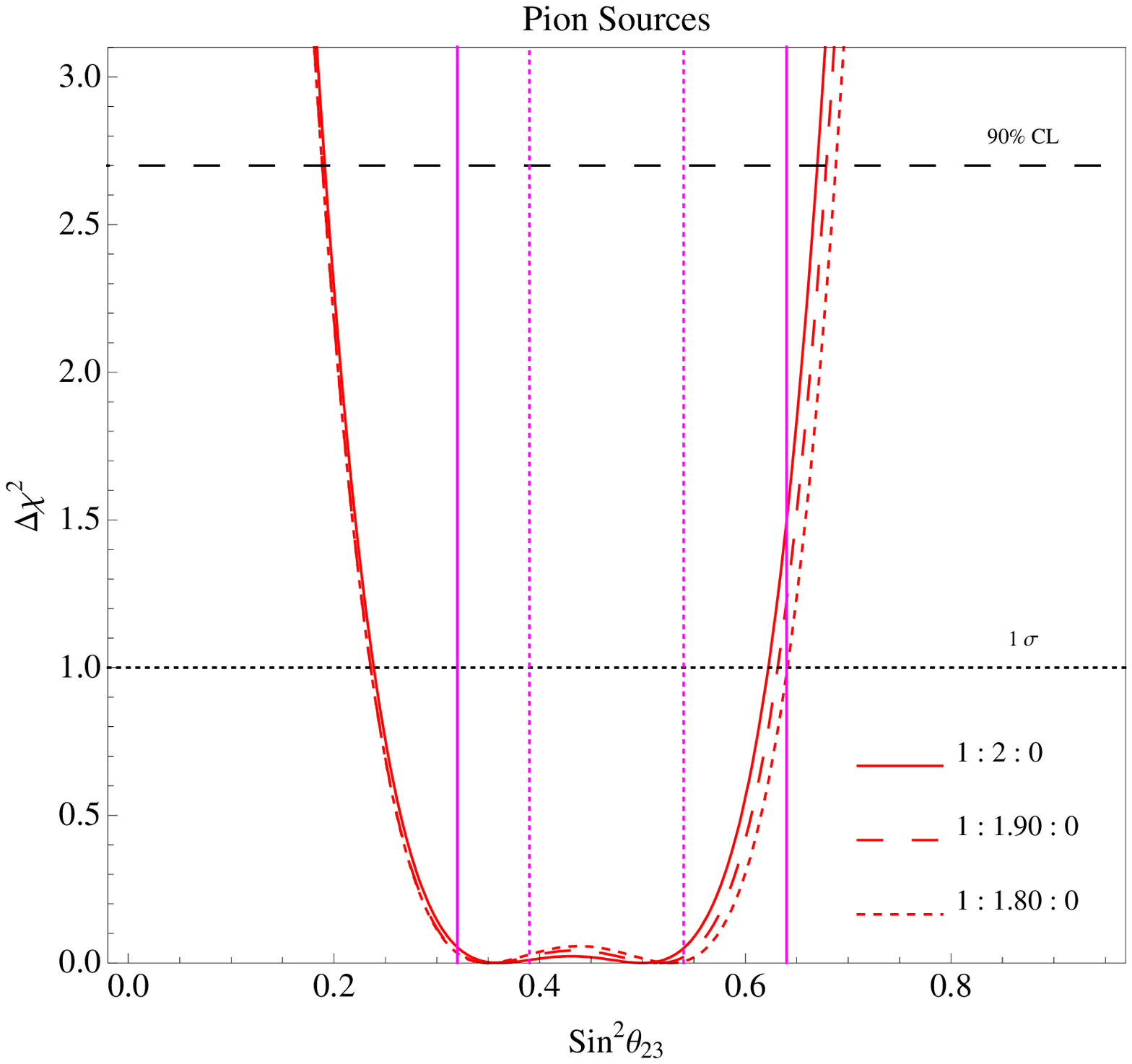} &
\includegraphics[width=8cm,height=6cm]{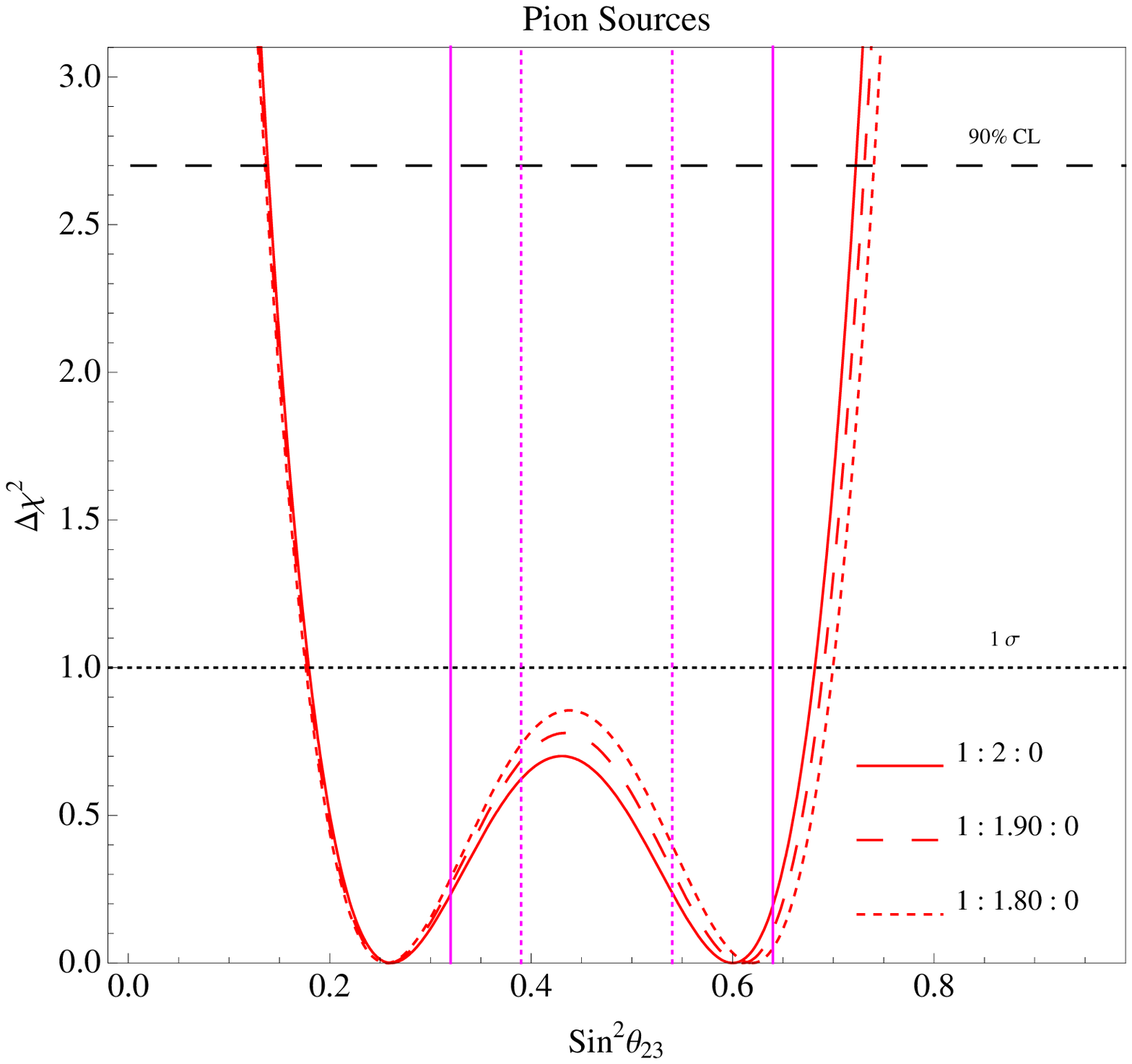}\\
\includegraphics[width=8cm,height=6cm]{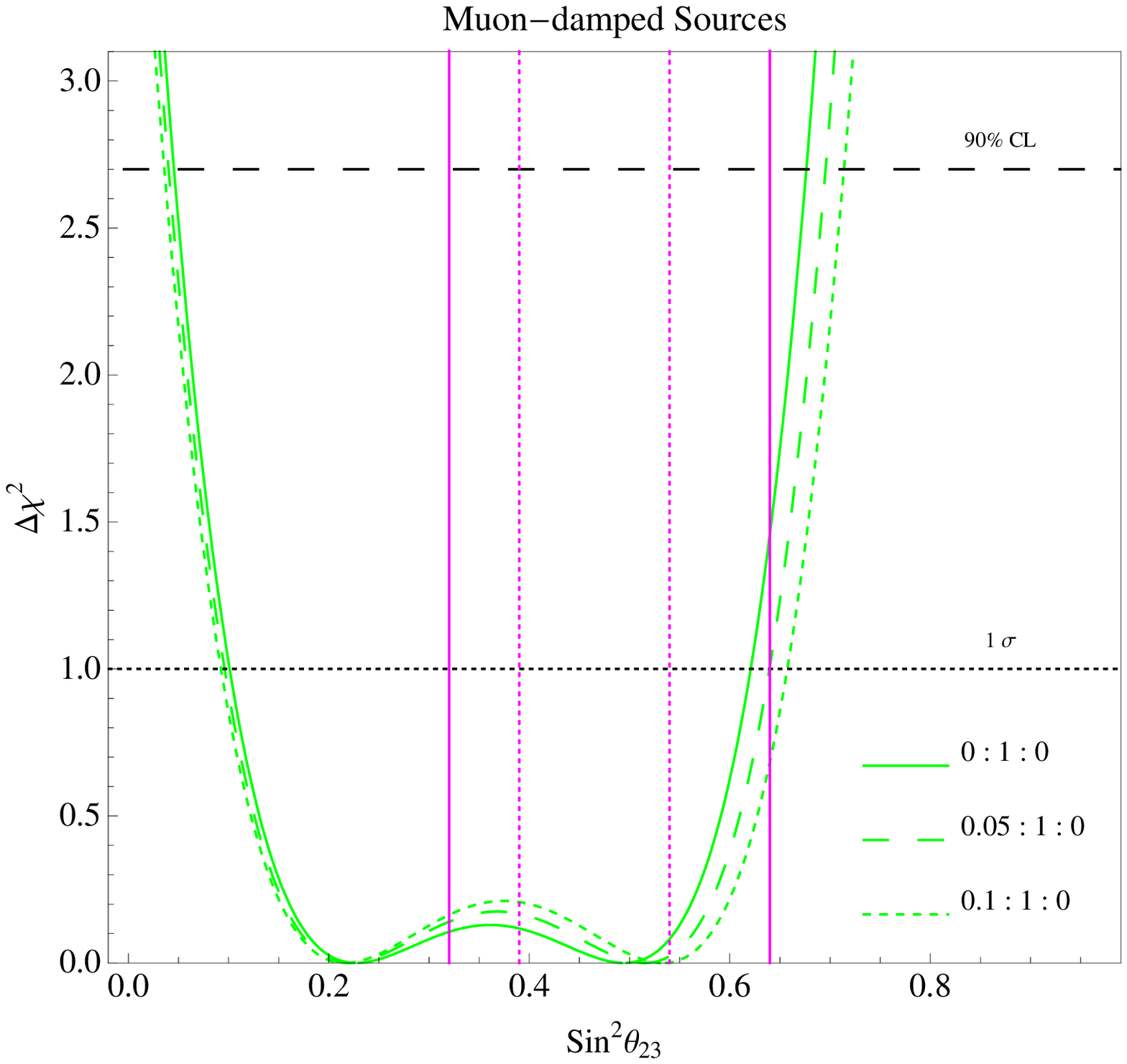} &
\includegraphics[width=8cm,height=6cm]{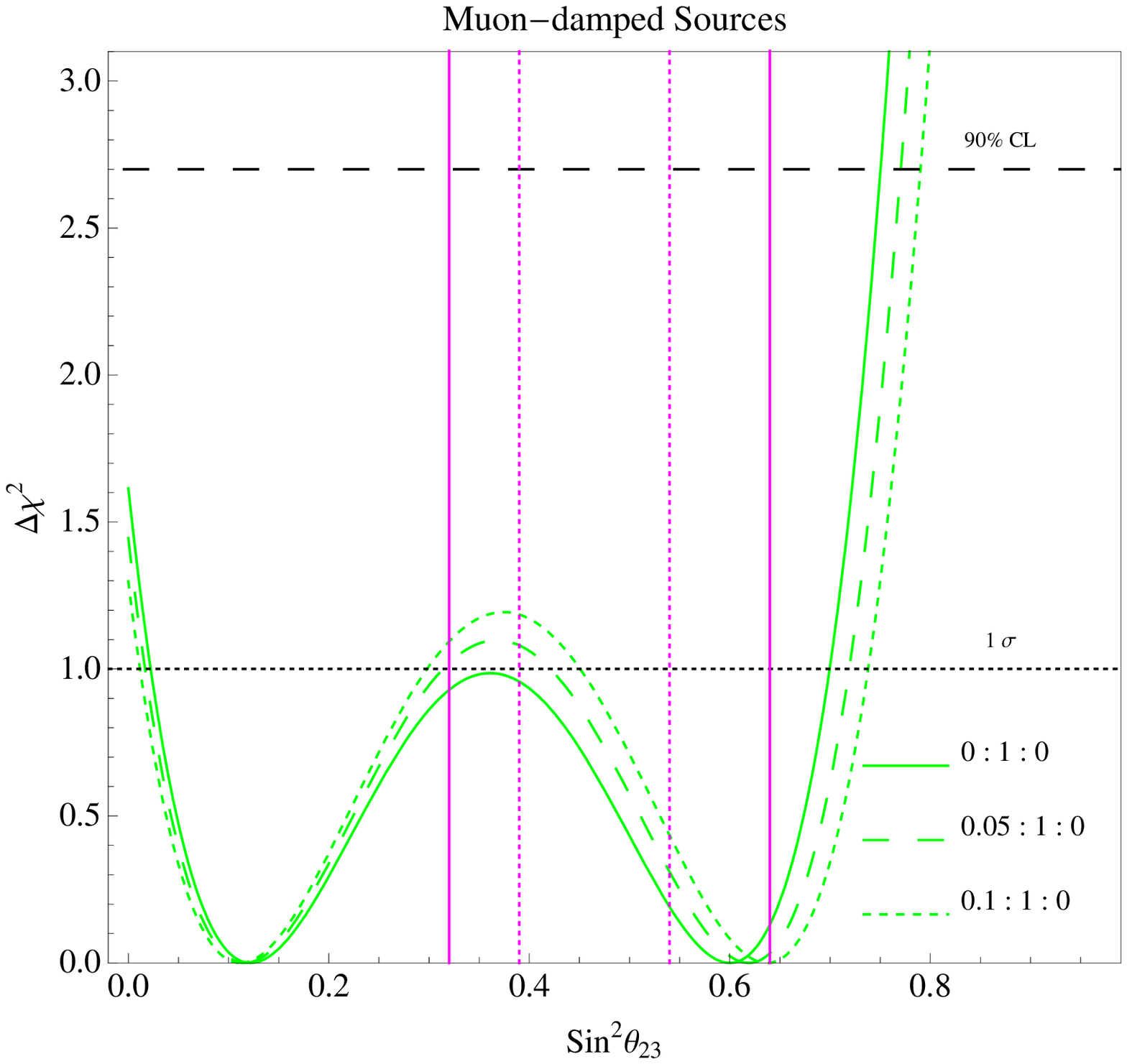}\\ 
\includegraphics[width=8cm,height=6cm]{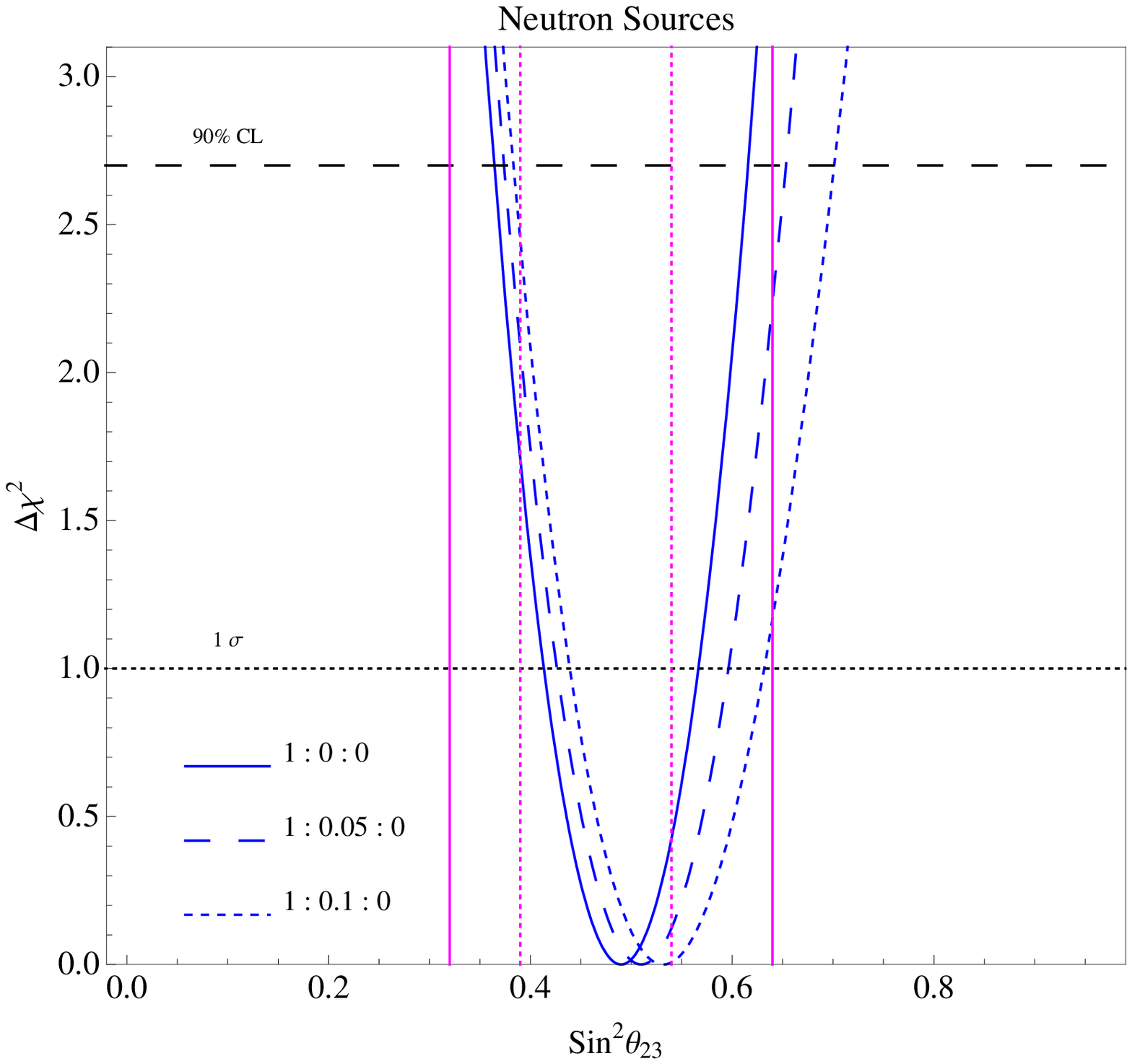} & 
\includegraphics[width=8cm,height=6cm]{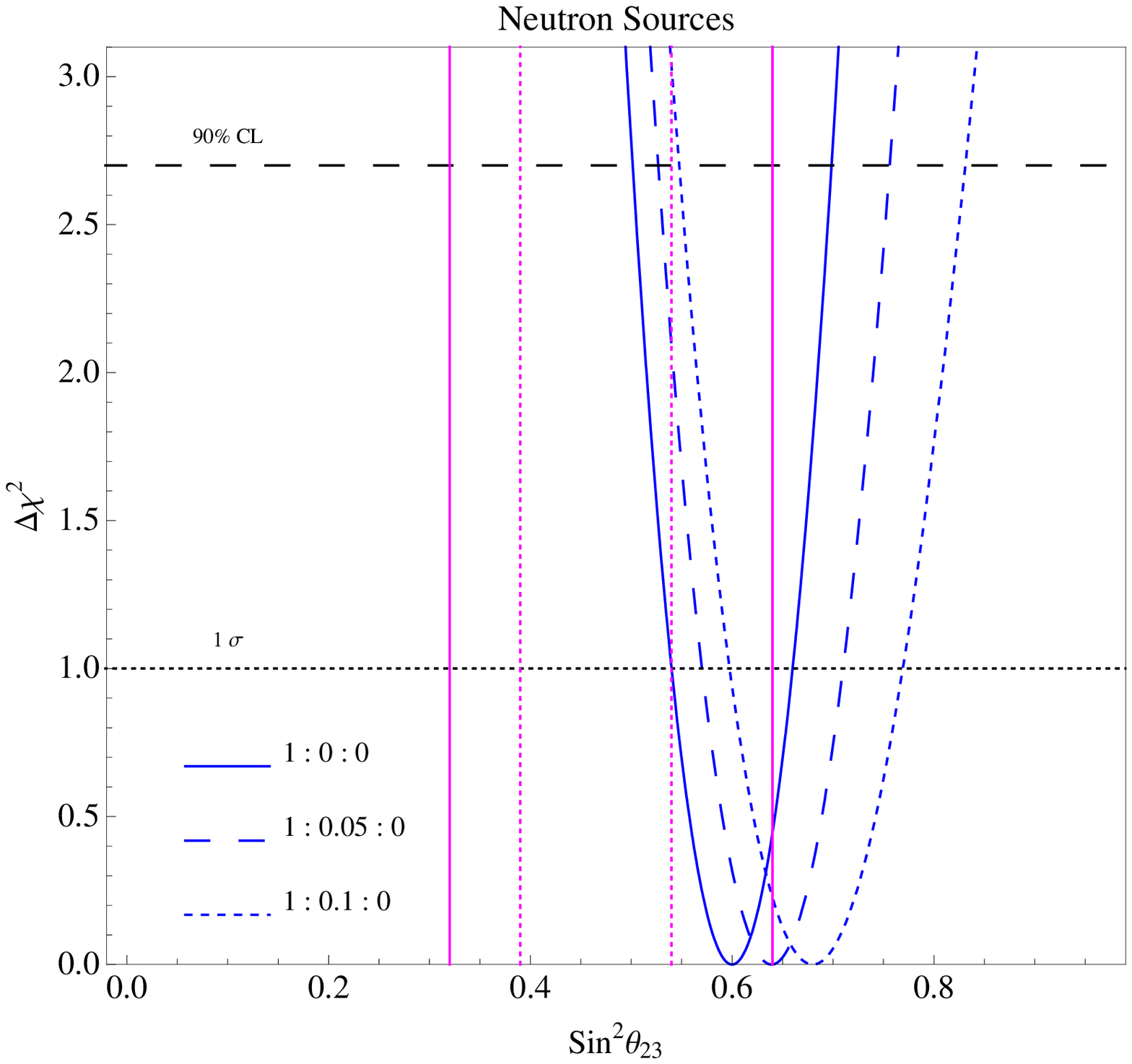}
\end{tabular}
\caption{Constraints on 
$\sin^2 \theta_{23}$ from a $\chi^2$ analysis obtained by 
measuring the 
ratio $T = \Phi_\mu/\Phi_{\rm tot}$ for the three different neutrino 
sources. 
The left panel assumes the ``tri-bimaximal values''
$T = \frac 13$, $\frac{7}{18}$ and $\frac 29$ as experimental values, 
while the right panel is for ``true values'' of the mixing 
angles given by Scenario 2. 
The current $ 1\sigma$ and 
$3\sigma$ ranges of $\sin^2 \theta_{23}$ 
are also indicated.}\label{fig:chi1} 
\end{figure}

We show the results of our $\chi^2$ fit using $T$ only 
in Fig.~\ref{fig:chi1}. The left panels show results for 
Scenario TBM while the right panels are for Scenario 2. The 
upper, middle and lower panels are for pion, muon-damped and 
neutron sources, respectively. In each panel the 
current $1\sigma$ and $3\sigma$ limits on $\sin^2\theta_{23}$ are given. 
We have allowed for impure sources by choosing initial 
flux compositions of  
$(1 : 2 : 0)$, $(0 : 1 : 0)$ and $(1 : 0 : 0)$, 
$(1 : 1.9 : 0)$, $(0.05 : 1 : 0)$ and 
$(1 : 0.05 : 0)$, as well as  
$(1 : 1.8 : 0)$, $(0.1 : 1 : 0)$ and $(1 : 0.1 : 0)$ for 
pion, muon-damped and neutron sources, respectively.  
Interestingly, impurity affects the  
neutron sources more strongly, and the 
impact is slightly larger on the ``dark-side'' of 
$\sin^{2}\theta_{23}$, i.e., for $\theta_{23} \ge \pi/4$. The 
``mexican hat'' shape for pion and muon-damped sources 
is easy to understand by looking at the 
$\sin^2 \theta_{23}$ dependence of $T$ in Fig.~\ref{fig:2d}. 
Therefore, for pion and muon-damped sources 
we get, in general, a two-fold 
degeneracy for every value of $T$. 
For Scenario 2 we 
get slightly larger values for $T$ and are hence 
no longer at the bottom 
of the parabolic shape. Therefore, the $\chi^2$ minima lie at 
lower and larger values of $\sin^{2}\theta_{23}$ and the two 
possible values of $\sin^2 \theta_{23}$ become further separated.

From Fig.~\ref{fig:chi1} we can infer that  
considering only the pion or muon-damped flavor ratio  
will bring limited information on $\sin^{2}\theta_{23}$. 
Indeed for these two sources and for Scenario TBM, 
we get an allowed $1\sigma$ range that is much 
bigger than the current $1\sigma$ 
range. With neutron sources, we get an 
allowed $1\sigma$ range for $\sin^{2}\theta_{23}$ from 0.42 to 0.64, 
which is a bit smaller than the current $3\sigma$ range.
For Scenario 2 (true value $\sin^2 \theta_{23} > 1/2$) we would not 
be able to corroborate it with either the pion or muon-damped 
sources, in particular because we obtain 
two allowed regions for $\sin^2\theta_{23}$. 
With neutron sources, and their linear dependence on $\sin^2\theta_{23}$, 
we expect to get only one allowed 
zone. The plot confirms this and in addition 
we see that in this case maximal mixing is ruled out by 
neutrino telescope data alone. 
To be more precise, from measuring $T$ with neutron sources we could 
constrain $\sin^{2}\theta_{23}$ to be greater than 0.54 
at the $1 \sigma$ level and greater than 0.5 at 90\% C.L. 

\begin{figure}[ht]
\begin{tabular}{cc}
\includegraphics[width=8cm,height=6cm]{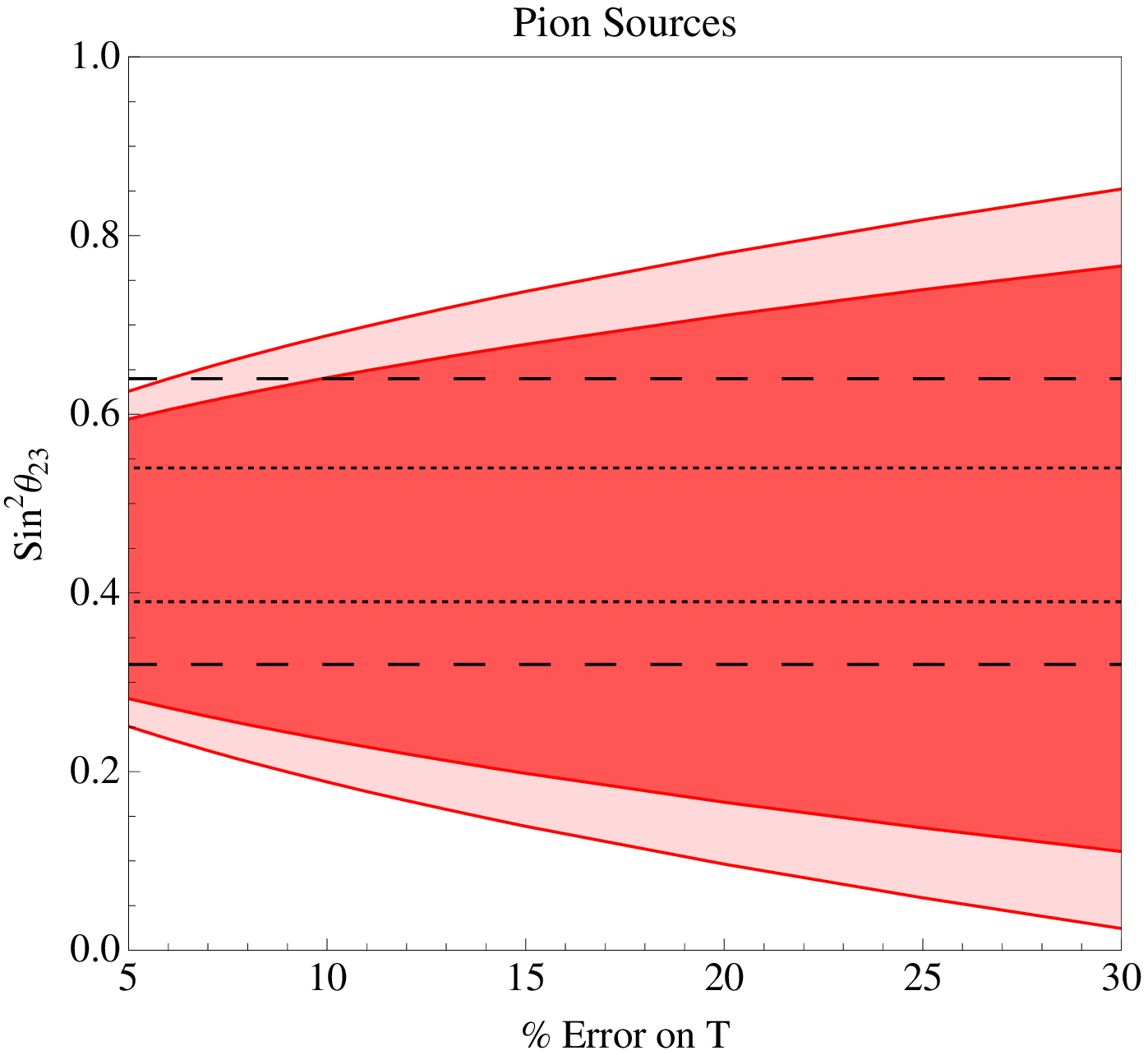} &
\includegraphics[width=8cm,height=6cm]{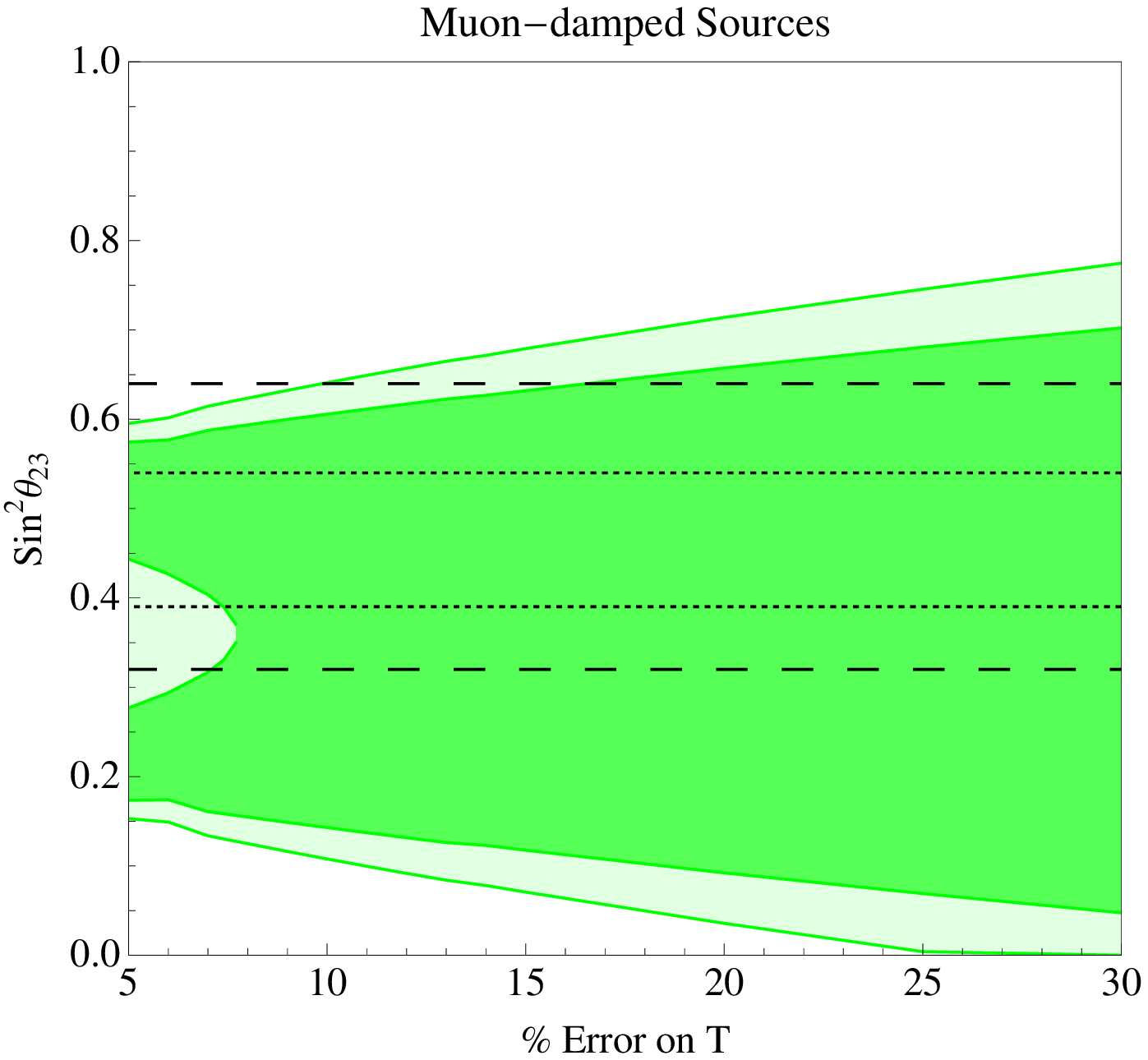} \\
\includegraphics[width=8cm,height=6cm]{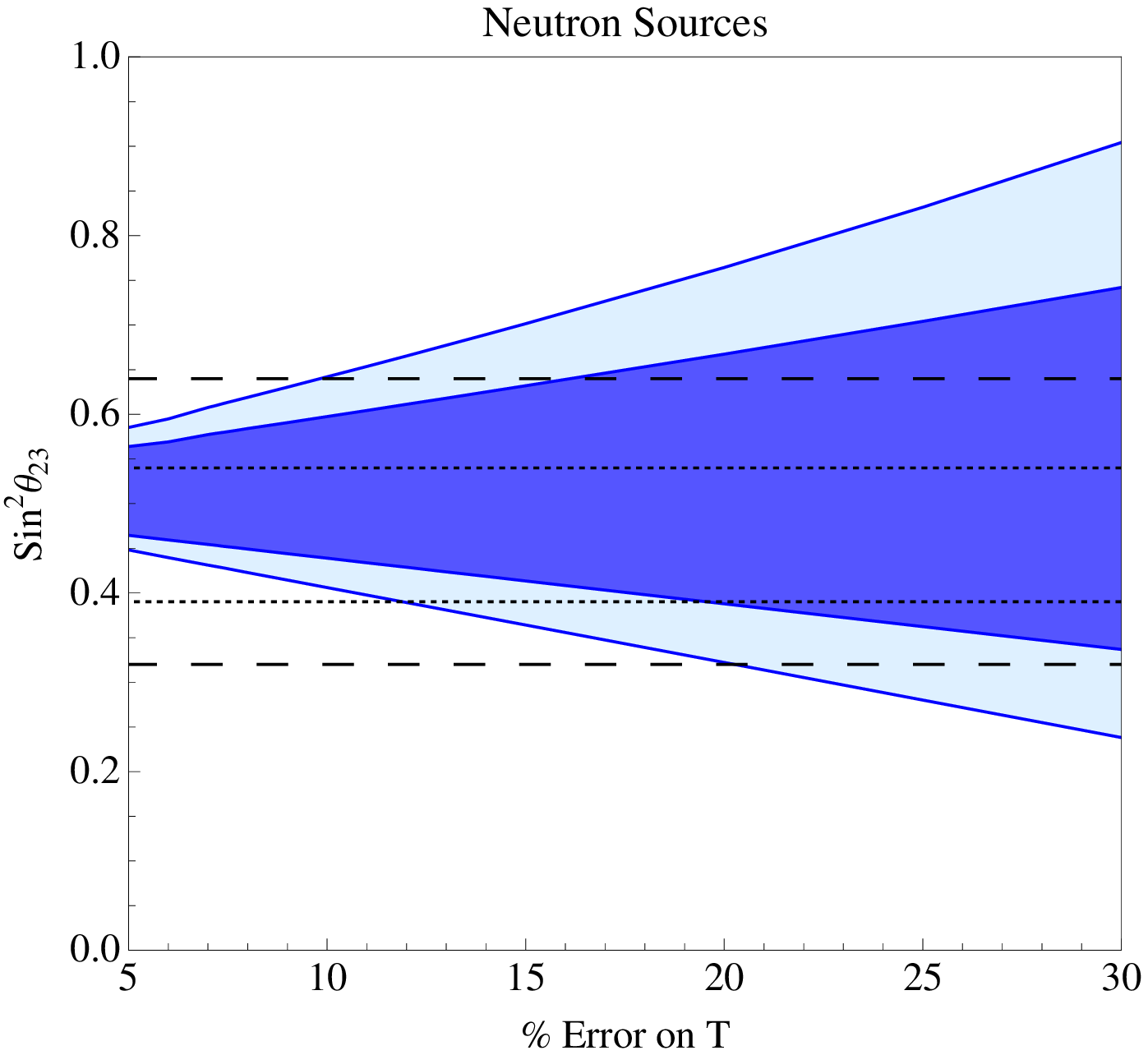} &
\includegraphics[width=8cm,height=6cm]{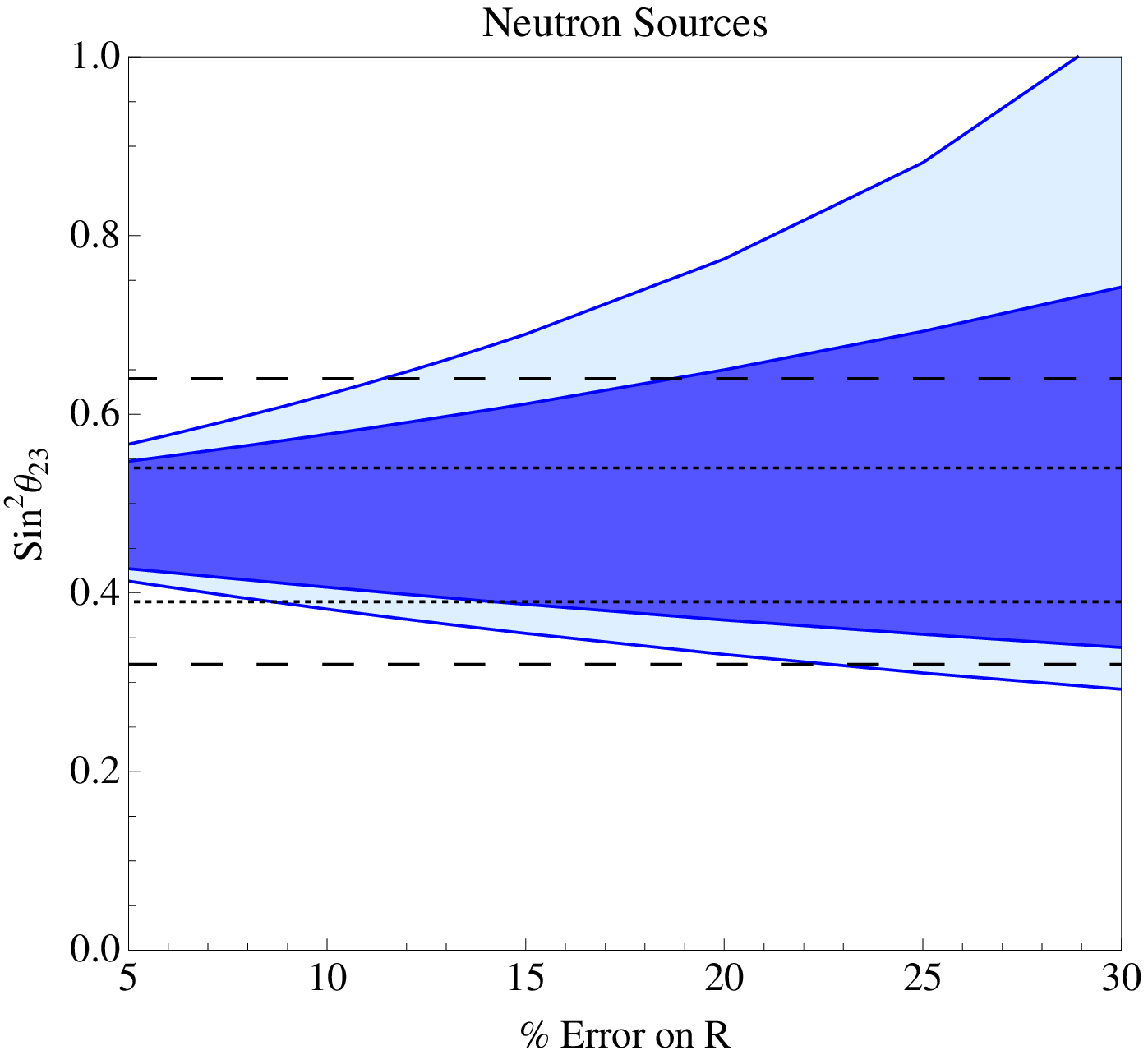} \\
\includegraphics[width=8cm,height=6cm]{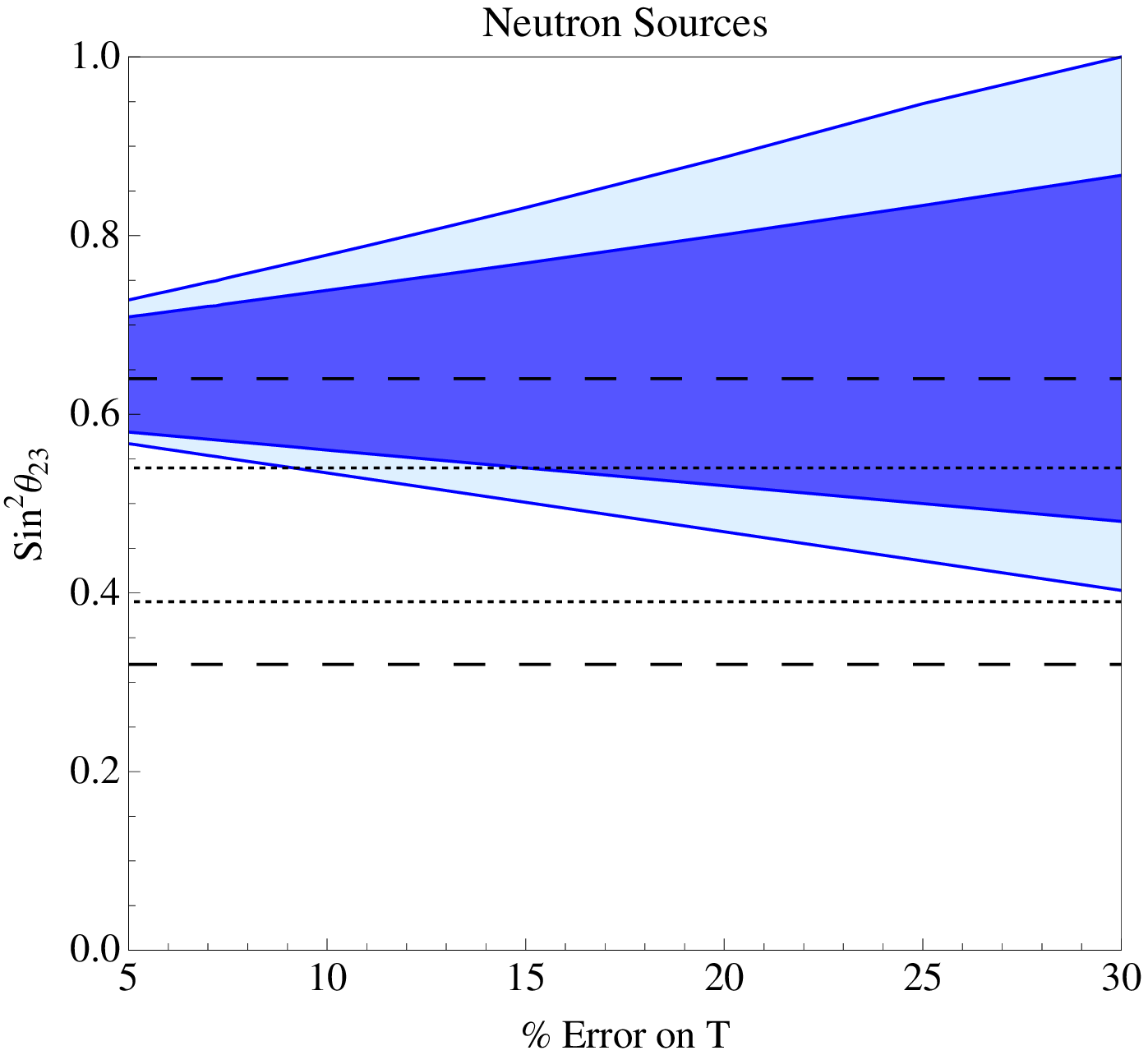} &
\includegraphics[width=8cm,height=6cm]{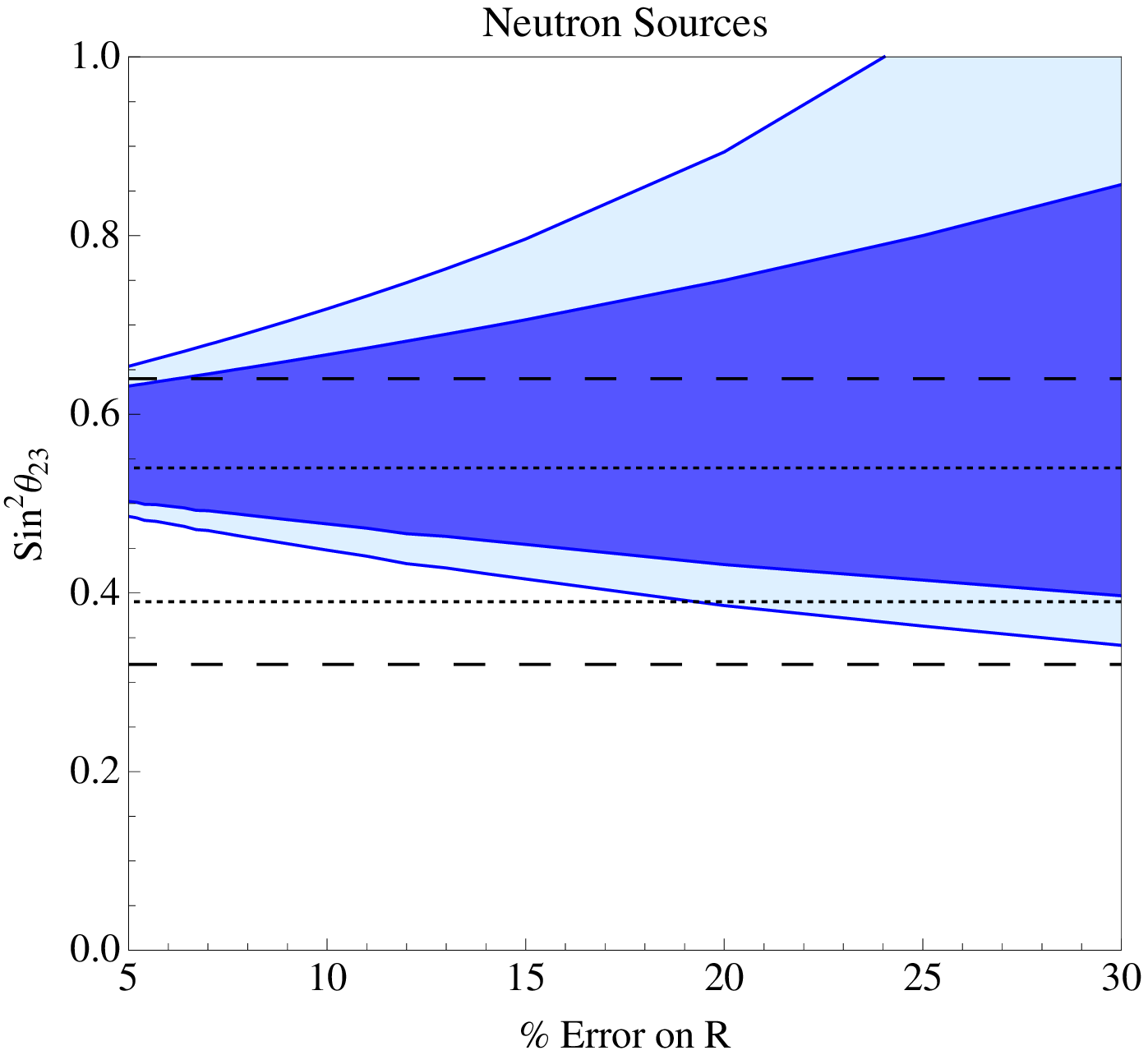}
\end{tabular}
\caption{Allowed range of $\sin^2 \theta_{23}$  
as a function of the error on the flavor ratio. The 
dark areas represent the $1 \sigma$ range, while the 
lighter areas are at 90\% C.L.  
The upper 4 plots are for scenario TBM, the 2 lower ones for 
scenario 2, where we only plot neutron sources. 
The current $1\sigma$ and 
$3\sigma$ ranges of $\sin^2 \theta_{23}$ 
are also indicated.}\label{fig:errorRatios} 
\end{figure}

Since these conclusions depend on the errors considered for 
the flavor ratios, 
we have generalized our analysis and studied the allowed range of 
$\sin^{2}\theta_{23}$ as a function of the error. 
In  Fig.~\ref{fig:errorRatios} we display the results. The 
upper 4 panels are for data generated for the TBM case while 
the lower 2 are for the neutron sources only and data corresponding
to Scenario 2. For TBM in data 
and with pion and muon-damped sources, we can infer that 
even for an extremely small error of the order of 5\%,   
the allowed range at 90\% C.L.~is bigger than the current ones. 
Neutron sources can impose an upper limit only for very small 
errors. A lower limit can be imposed if the error on the 
flavor ratio is smaller than 18\%. If Scenario 2 was true, 
then neutron sources could rule out maximal mixing and 
give information on the 
octant of $\theta_{23}$, and Fig.~\ref{fig:errorRatios} shows that 
the $T$ ratio is better suited than $R$ for this purpose. 

\begin{figure}[ht]
\begin{tabular}{cc}
\includegraphics[width=8cm,height=6cm]{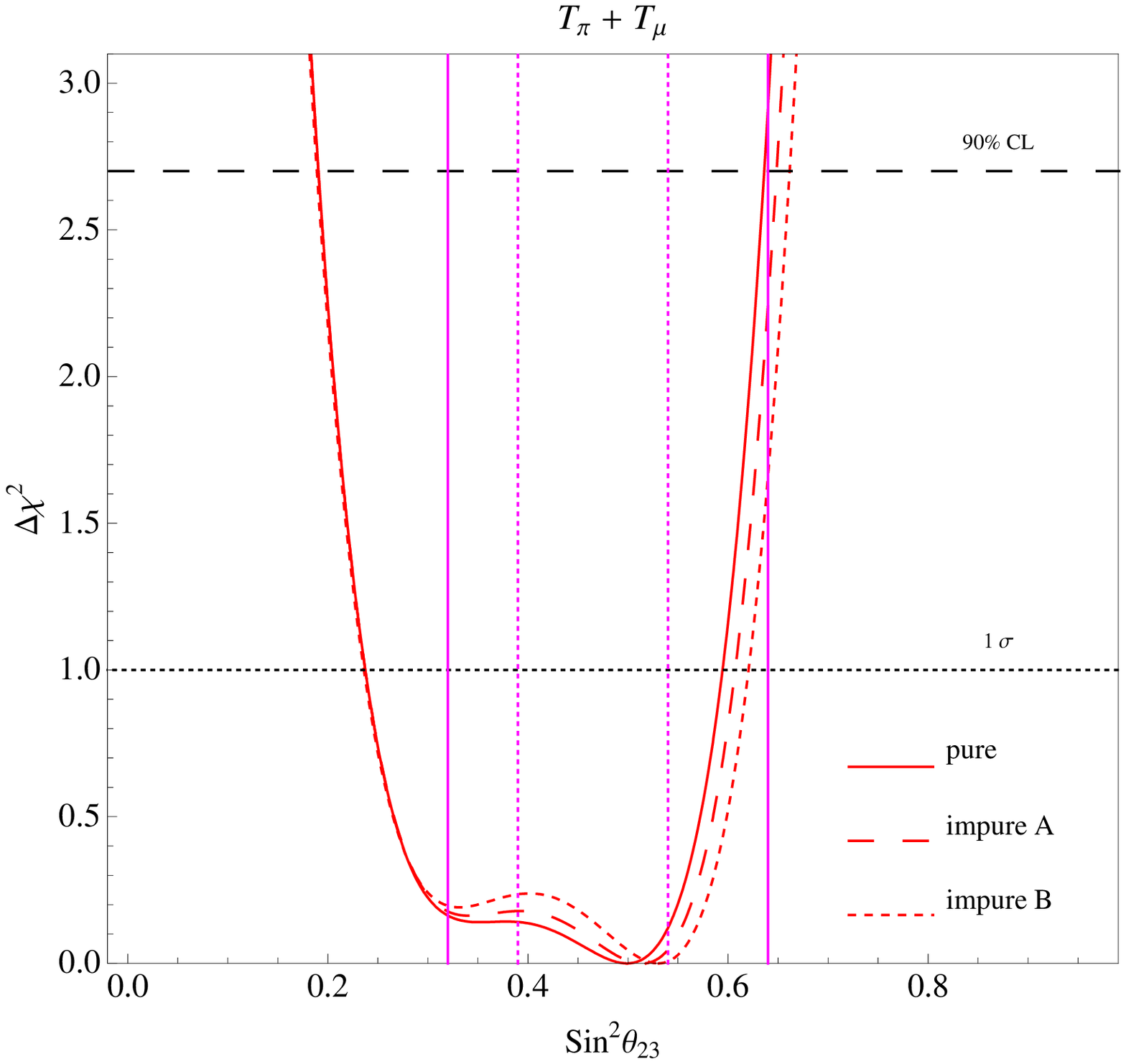} & 
\includegraphics[width=8cm,height=6cm]{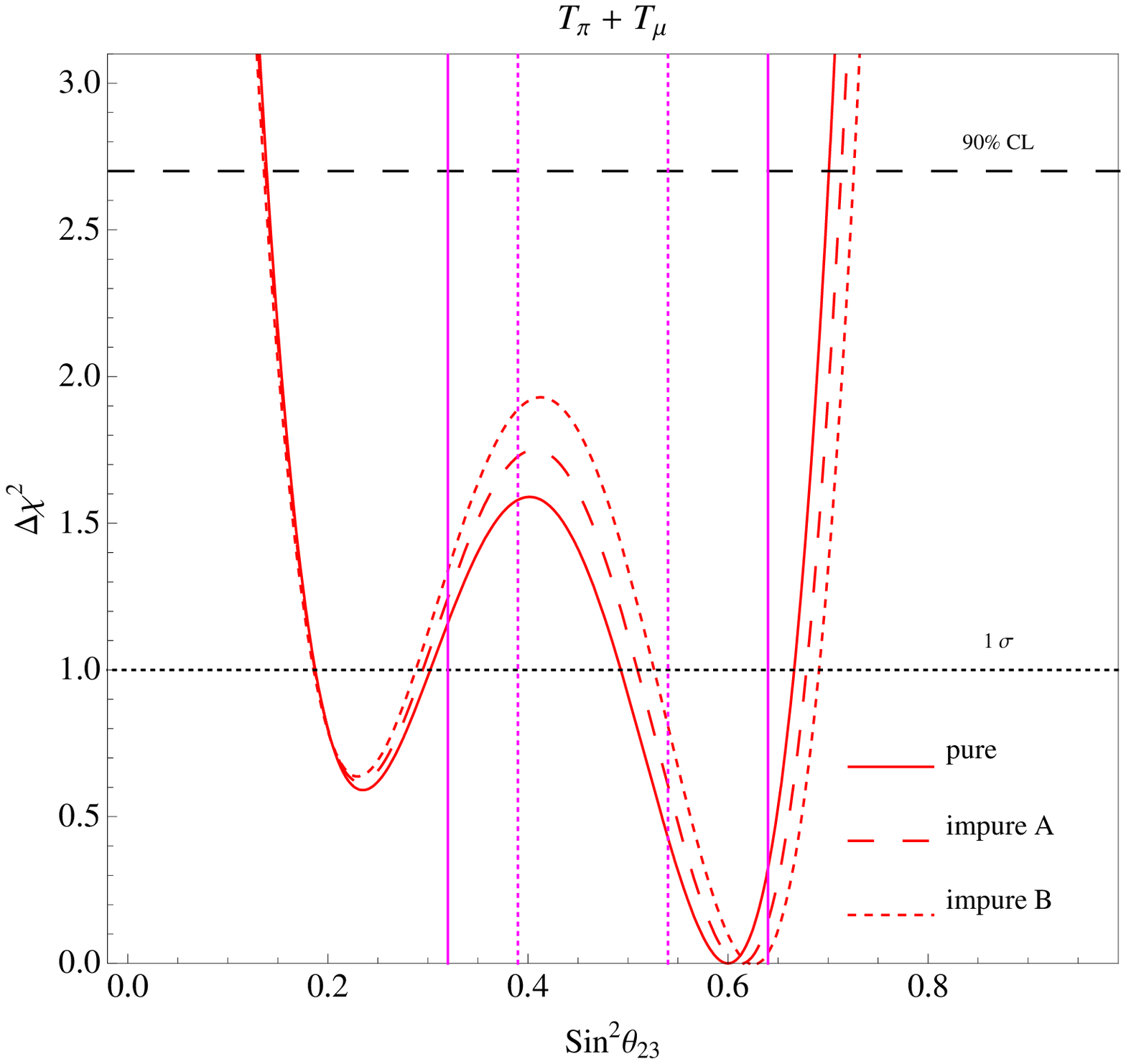} \\ 
\includegraphics[width=8cm,height=6cm]{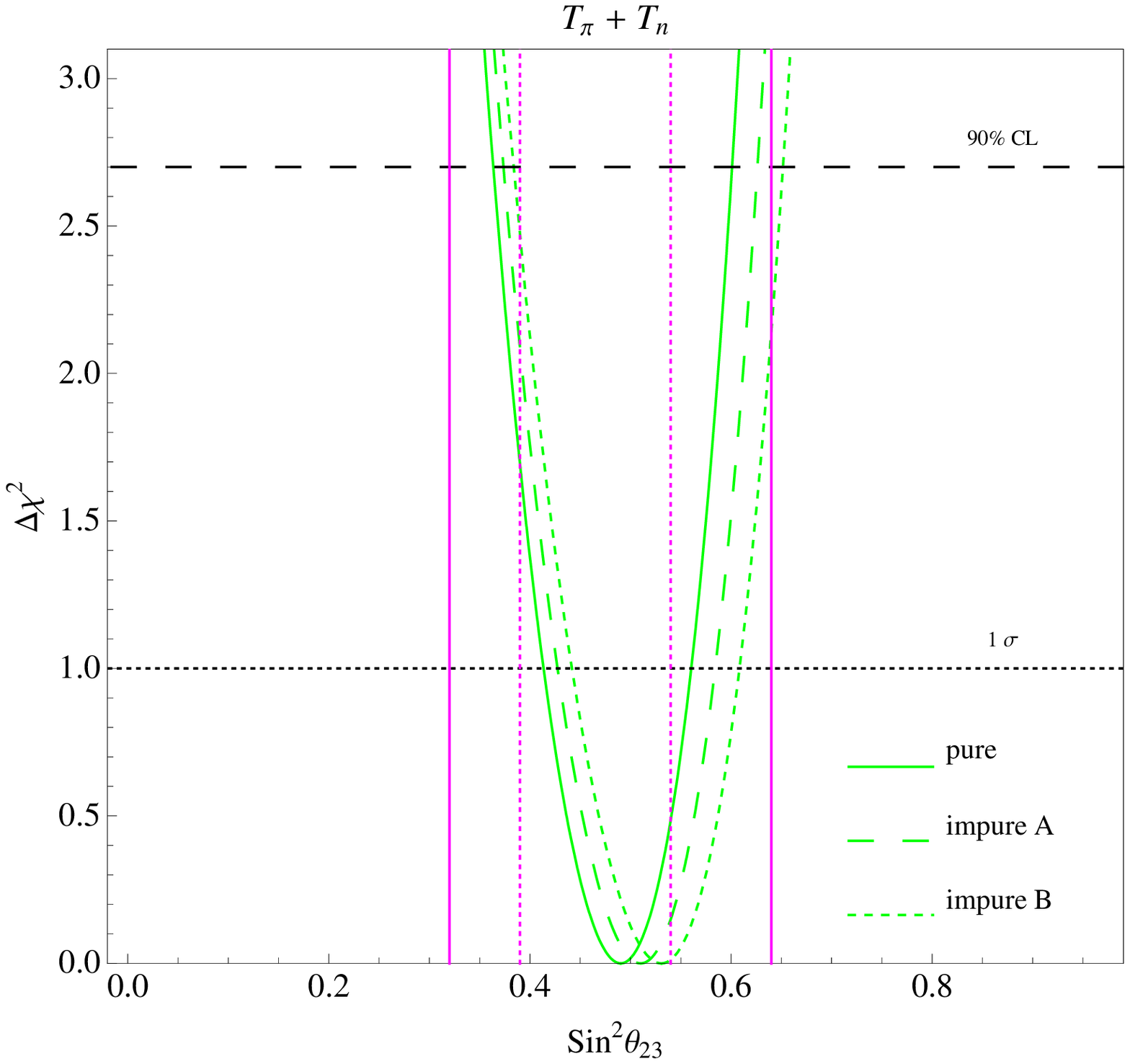} & 
\includegraphics[width=8cm,height=6cm]{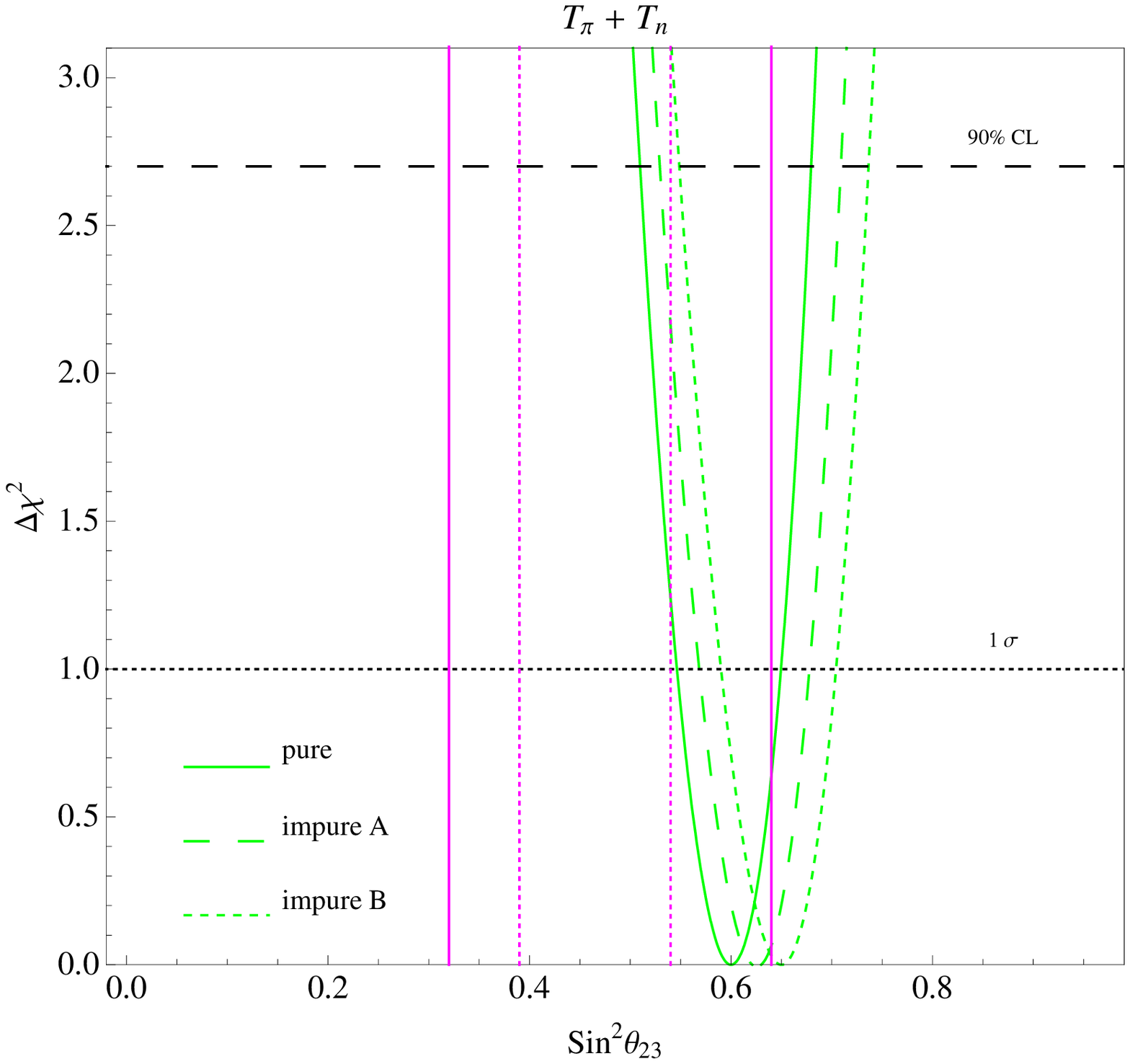}\\ 
\includegraphics[width=8cm,height=6cm]{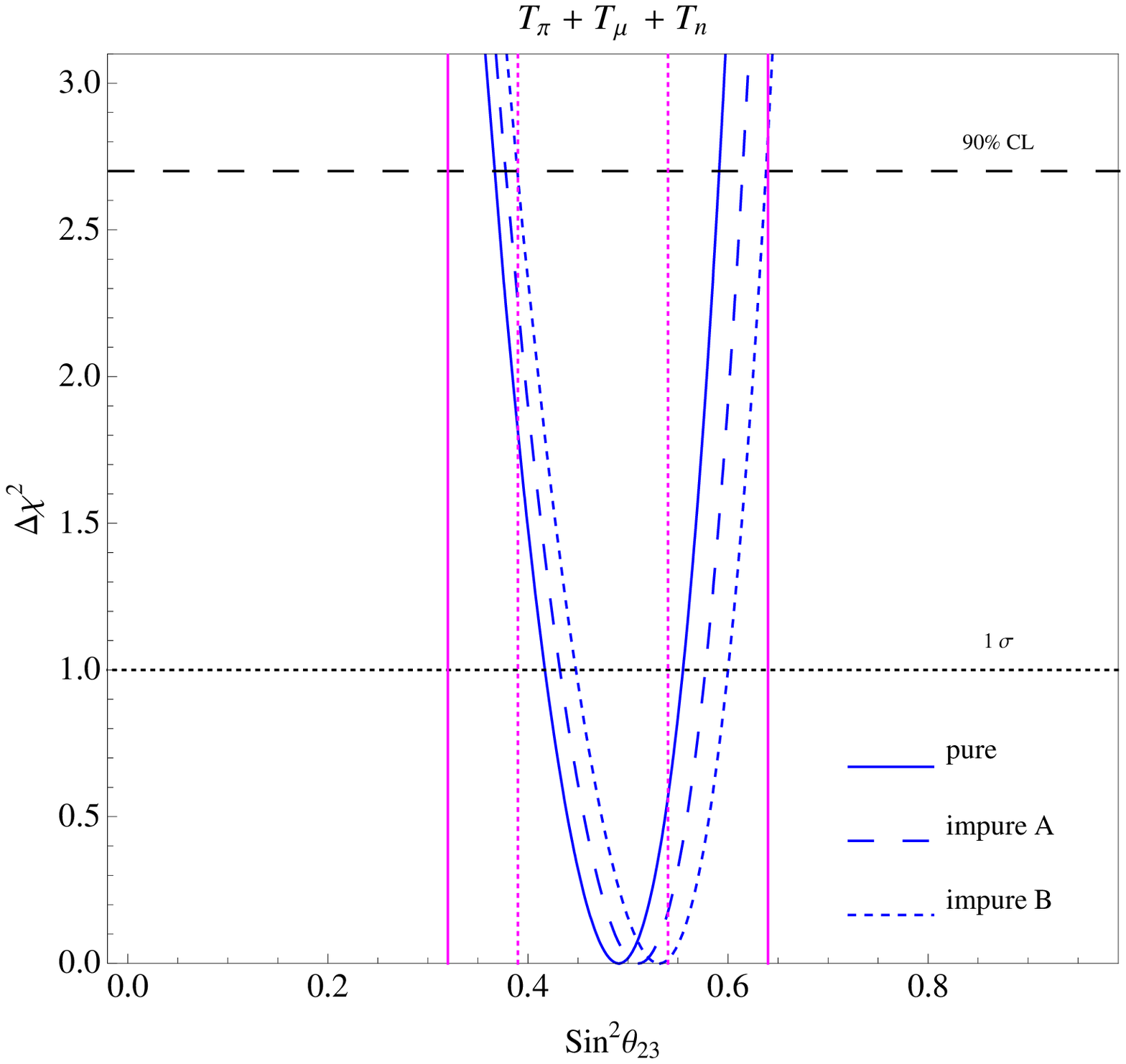} & 
\includegraphics[width=8cm,height=6cm]{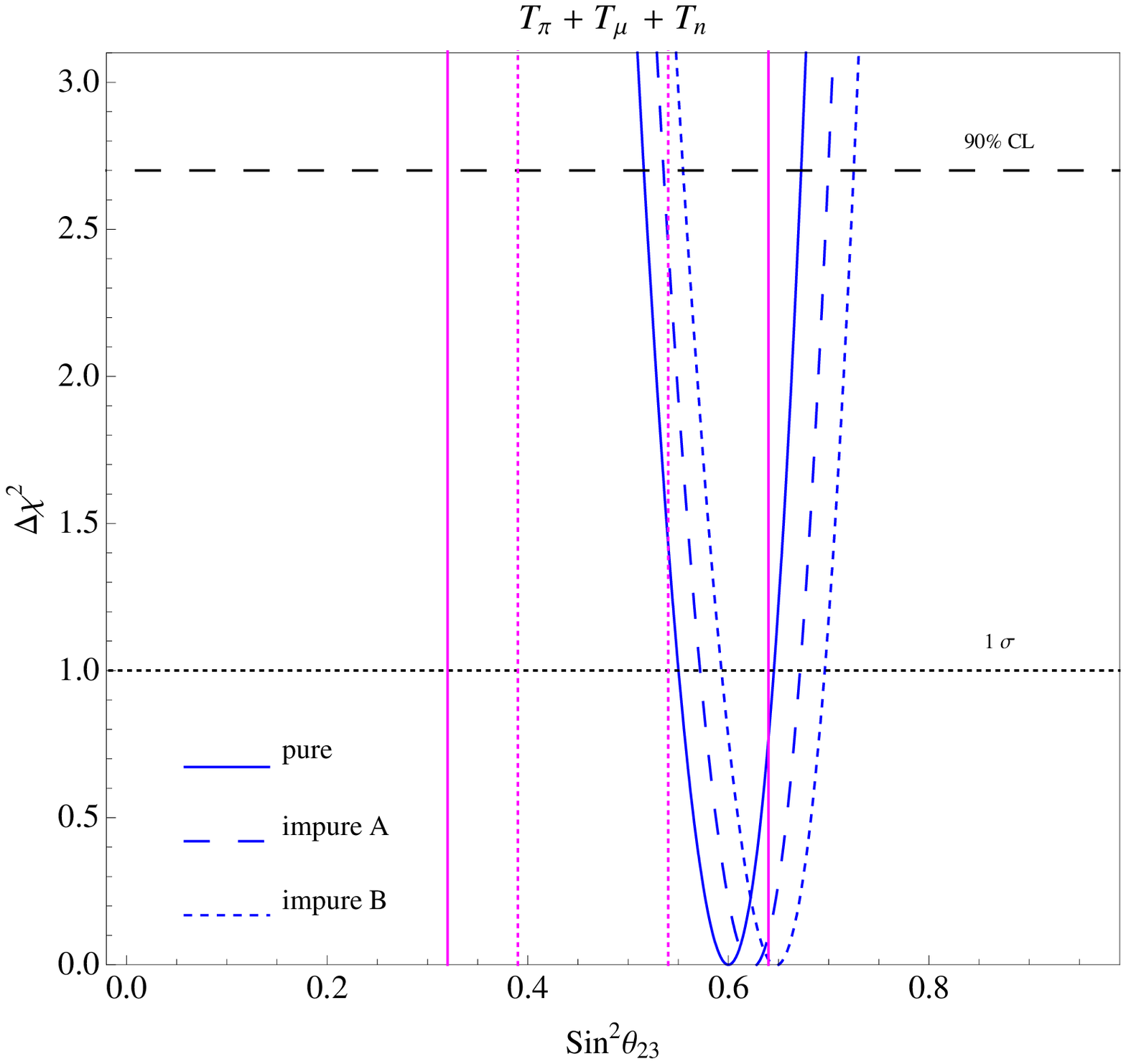} 
\end{tabular}
\caption{Same as Fig.~\ref{fig:chi1}, but for 
combined measurements of $T$ from different sources. With ``pure'' 
we refer to the initial flavor composition $(1 : 2 : 0)$, $(0 : 1 : 0)$ 
and $(1 : 0 : 0)$, with ``impure A'' 
we refer to $(1 : 1.9 : 0)$, $(0.05 : 1 : 0)$ and $(1 : 0.05 : 0)$, 
while for ``impure B'' 
we refer to $(1 : 1.8 : 0)$, $(0.1 : 1 : 0)$ and $(1 : 0.1 : 0)$.}
\label{fig:chi2}
\end{figure}

To improve the sensitivity of neutrino telescopes to $\theta_{23}$, 
one should combine measurements of $T$ from different sources. 
In Fig.~\ref{fig:chi2} we have considered the 
combination of pion and muon-damped sources (upper panels), 
of pion and neutron sources (middle panels) 
and of all the three different 
type of sources (lower panels).  
The $\chi^{2}$ function is, for example, 
\begin{equation} \label{eq:chi22}
\chi^{2} = 
\left(\frac{(T_{\pi})_{\rm th} - (T_\pi)_{\rm exp}} 
{\sigma_{(T_\pi)_{\rm exp}}}\right)^{2} + 
\left(\frac{(T_{\mu})_{\rm th} - (T_\mu)_{\rm exp}} 
{\sigma_{(T_\mu)_{\rm exp}}}\right)^{2} 
+ 
\sum_{ij=12,13} 
\left(\frac{s^{2}_{ij}-(s^{2}_{ij})_{\text{best-fit}}}
{\sigma_{s^{2}_{ij}}}\right)^{2} 
\end{equation}
when measurements from pion and muon-damped sources are combined. 
The left panels show results for the Scenario TBM, while 
the right panels showcase Scenario 2. 
It should be noted that the two-fold degeneracy 
with respect to $\sin^2 \theta_{23}$ is hardly present anymore 
when two or more $T$ ratios are combined. 
Three different cases are plotted in the figures. With ``pure'' 
we refer to initial flavor compositions $(1 : 2 : 0)$, $(0 : 1 : 0)$ and 
$(1 : 0 : 0)$, with ``impure A'' 
we refer to $(1 : 1.9 : 0)$, $(0.05 : 1 : 0)$ and $(1 : 0.05 : 0)$, 
while for ``impure B'' 
we refer to $(1 : 1.8 : 0)$, $(0.1 : 1 : 0)$ and $(1 : 0.1 : 0)$. 
Again, the effect of impurity is stronger for $\sin^2 \theta_{23} > 1/2$. 
In general, impurity shifts the $\chi^2$ minimum to larger values 
of $\sin^2 \theta_{23}$. Analyzing the plots, we can conclude that 
the combinations of pion and muon-damped sources will not 
give us precise 
information on $\sin^{2}\theta_{23}$ and that only the combination 
with neutron sources improves substantially the sensitivity. Indeed,
we can notice that the combined 
constraint from pion and neutron sources looks 
basically identical to the result of 
neutron sources alone (see Fig.~\ref{fig:chi1}). Moreover, 
if all three sources are combined, i.e., if muon-damped 
sources are added, hardly any improvement is seen. 
For Scenario 2, 
we could determine the octant of $\theta_{23}$ 
at the $1 \sigma$ level in case we combine $T_{\pi}$ and 
$T_{n}$ or all the three flavor ratios. We can, therefore, 
conclude that the importance of the neutron source is crucial to 
provide a good sensitivity to $\sin^{2}\theta_{23}$. 

\begin{figure}[ht]
\begin{tabular}{cc}
\includegraphics[width=8cm,height=6cm]{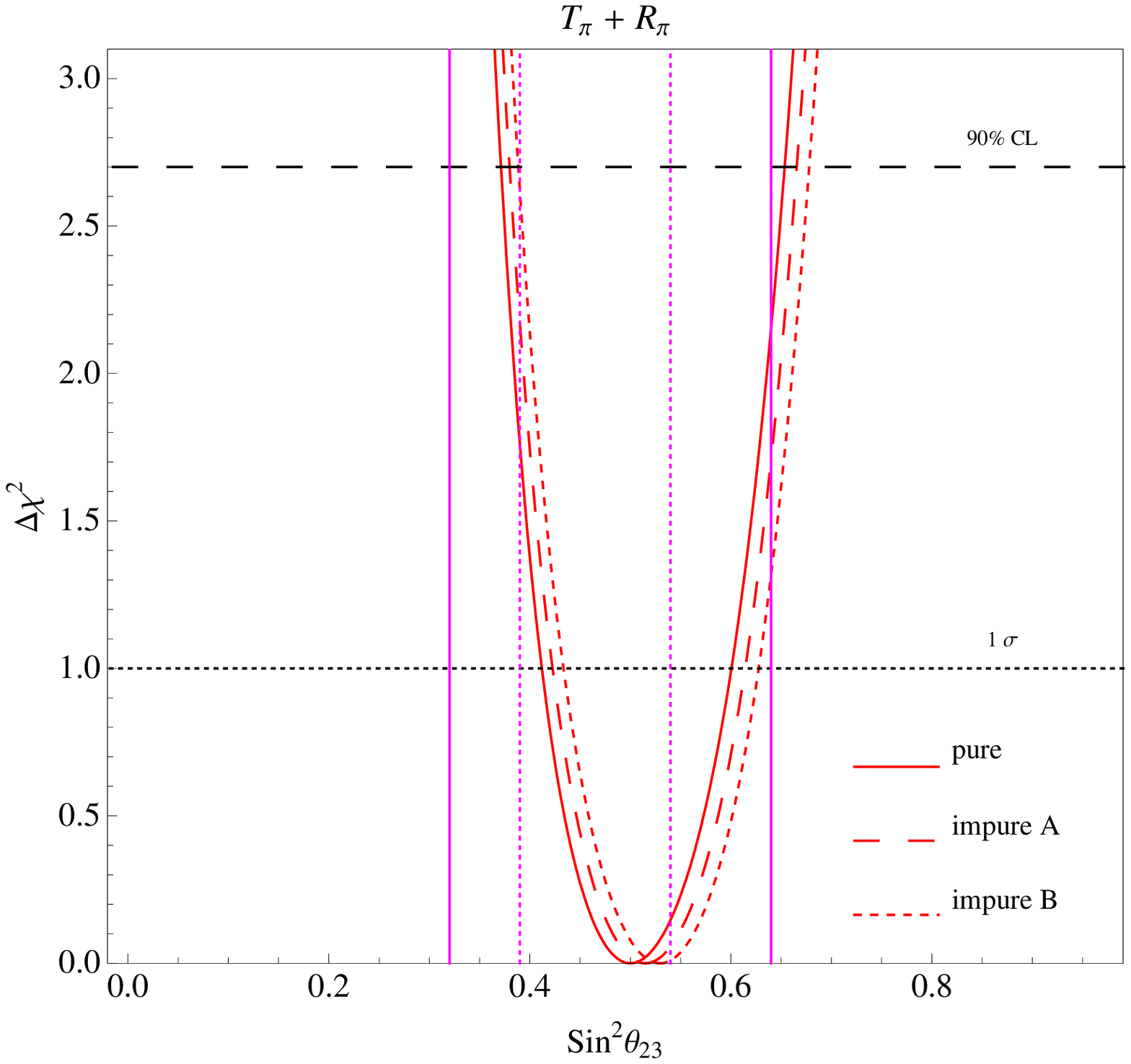} & 
\includegraphics[width=8cm,height=6cm]{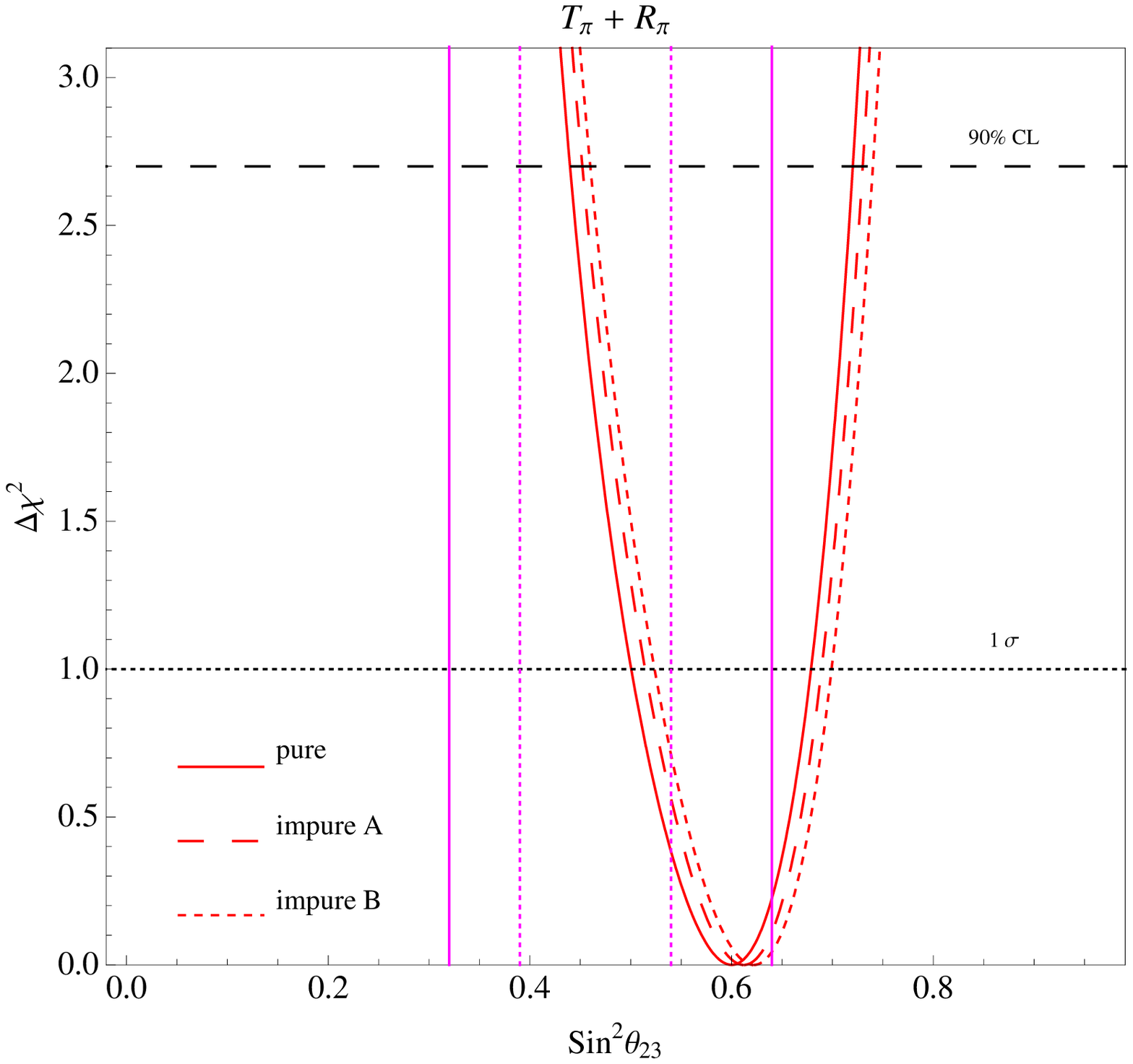} \\ 
\includegraphics[width=8cm,height=6cm]{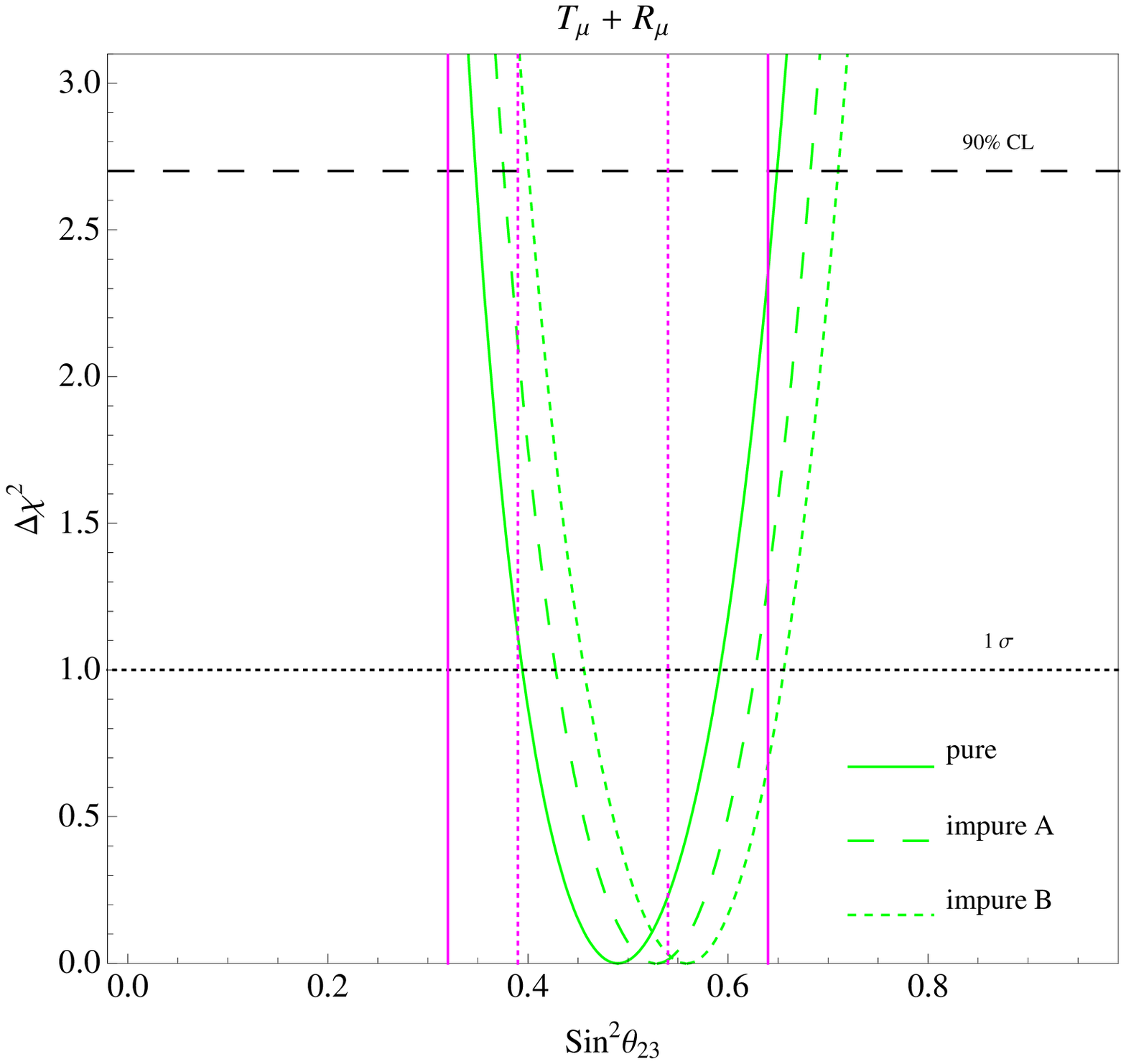} & 
\includegraphics[width=8cm,height=6cm]{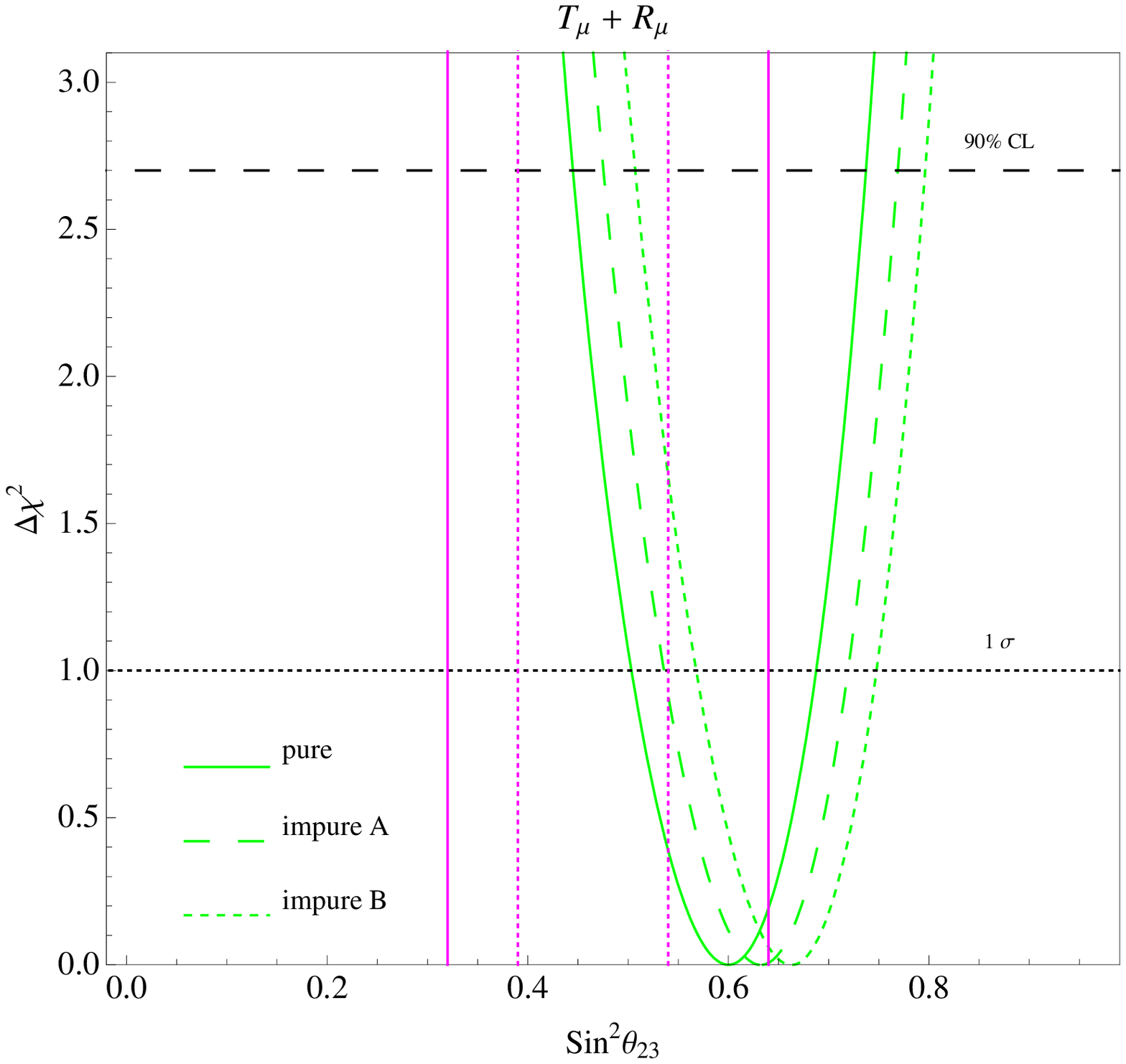} \\
\includegraphics[width=8cm,height=6cm]{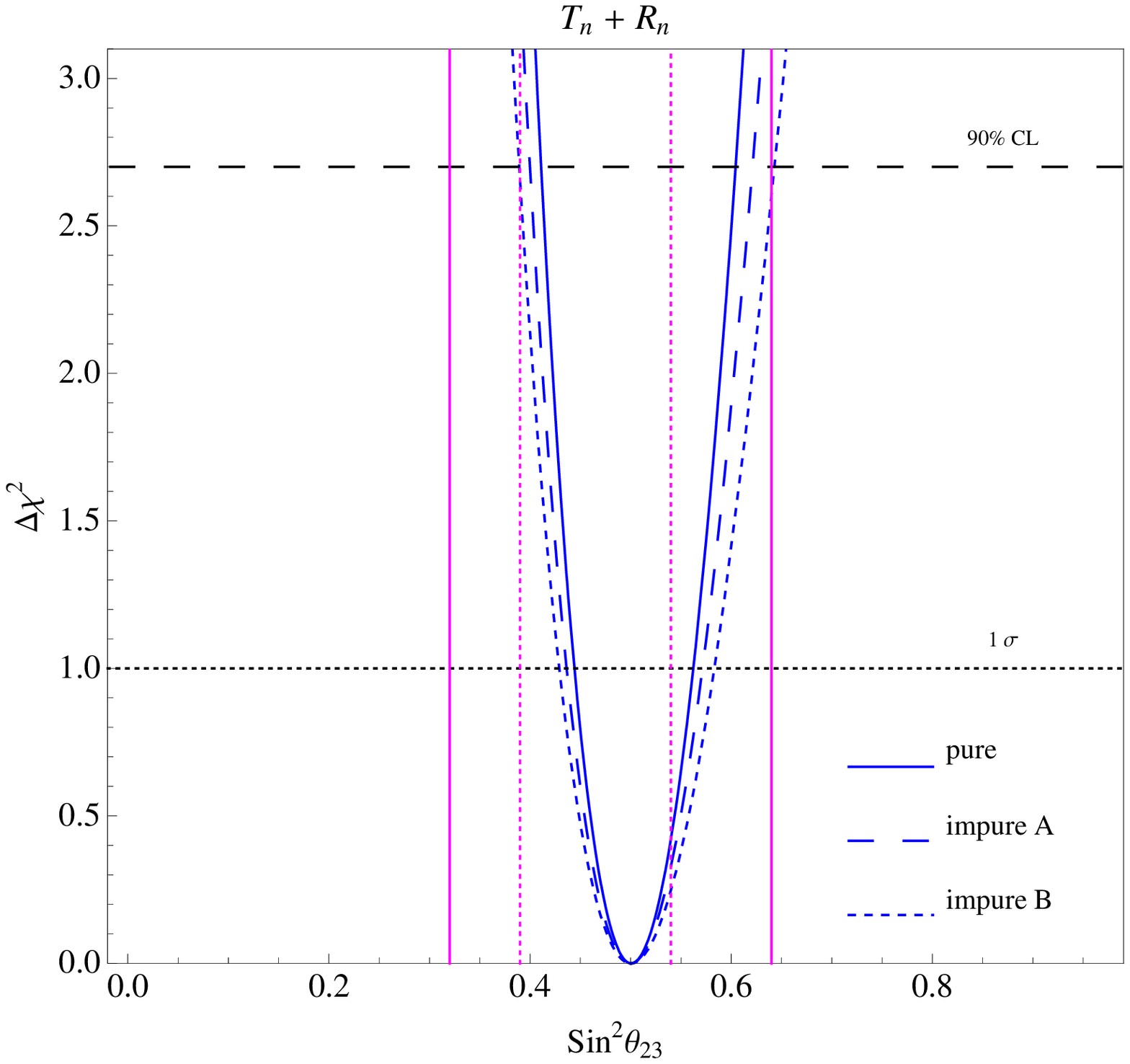} & 
\includegraphics[width=8cm,height=6cm]{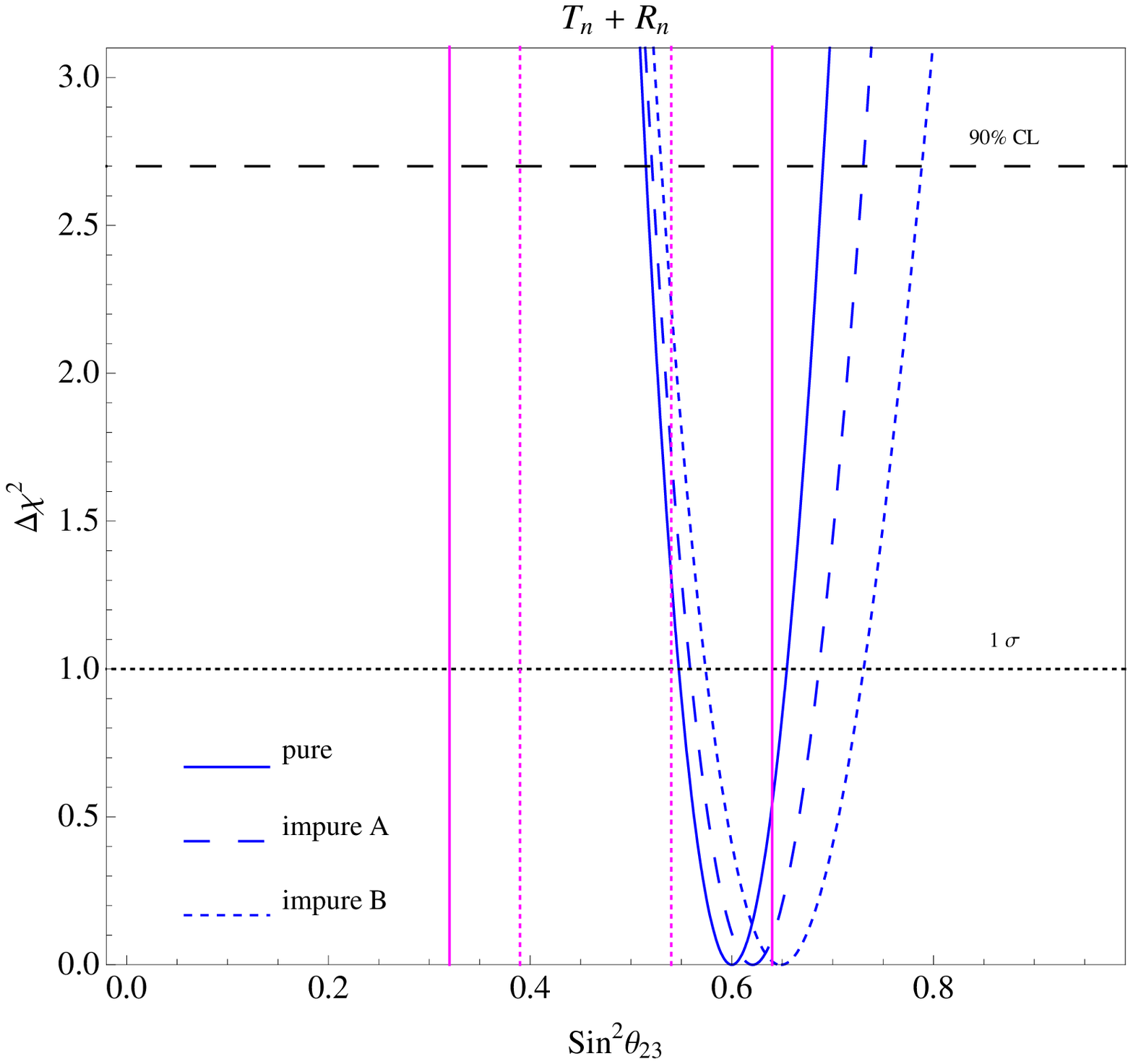} 
\end{tabular}
\caption{Same as Fig.~\ref{fig:chi1}, but for combined measurements 
of $T$ and $R$ ratios. 
}\label{fig:chi3}
\end{figure}

Better constraints on $\theta_{23}$ are obtained when the 
ratios $T$ and $R$ are combined. The $\chi^2$ function 
is now the same as in Eq.~(\ref{eq:chi21}) with an 
appropriate term including $R$ added. The result of the minimization  
is shown in Fig.~\ref{fig:chi3}, and the improvement with 
respect to the other plots is obvious. The effect of impurity is 
however stronger. This is understandable from 
the approximate expressions 
of $T$ and $R$ in Eqs.~(\ref{eq:T},\,\ref{eq:R}), in which 
the impurity factors $\zeta$ and $\eta$ have more sizable 
prefactors in $R$. 
 
To summarize, once one combines $T$ and $R$, 
the $\chi^2$ curves become less broad, but the 
shift due to impurity becomes more sizable. 
The best constraint is obtained by combining $T$ and $R$ 
for neutron sources. 
Indeed, Fig.~\ref{fig:chi3} shows that the $1\sigma$ range 
for $\sin^{2}\theta_{23}$ is of the same order as the current one. 
For Scenario 2 we can exclude maximal mixing  
at roughly $90 \%$ C.L.~level. This conclusion 
depends also on the error considered for $R$. Varying it we show 
in Fig.~\ref{fig:errorRatios} 
the allowed range of $\sin^{2}\theta_{23}$ if only $R$ was measured. 

The most optimistic scenario would occur when both ratios can 
be measured for all three sources. We checked that the outcome 
is basically identical to combining $T$ and $R$ 
for neutron sources, showing once again that these sources are best 
suited for $\theta_{23}$ constraints.

We stress that 
throughout this study we have tried to extract information on 
only the mixing angle $\theta_{23}$ and  
assumed that the kind of astrophysical 
source is known. 
Note that the values for the $T$ 
and $R$ ratios of different sources are well separated even if the  
errors on the flux ratios are bigger than the ones we are considering. 
In addition, the neutron beams, for which the sensitivity to 
$\sin^{2}\theta_{23}$ is stronger, have the advantage that the ratio $T$ 
is rather small, and $R$ is very large. This will allow more easily 
to distinguish one source from the others.
A detailed analysis on the problematics related to the 
determination of the type of source is beyond the task of this work and 
will be published separately.

\section{\label{sec:concl}Summary and Conclusions}

Neutrino mixing affects the flux ratios of ultra high energy neutrinos 
arriving on earth. The exact values of the ratios are determined by  
the values of the mixing parameters. A series of papers have looked 
into the impact of standard and non-standard properties of 
neutrinos on the flux ratios. In this paper we argued that 
within the framework of the standard picture with three 
stable neutrinos, the mixing angle $\theta_{23}$ has the 
maximum effect on the flux ratios. We performed a statistical 
test to ascertain quantitatively the extent to which 
this mixing angle can be constrained by measuring the flux 
ratios of ultra high energy neutrinos. 

We defined two kind of flux ratios, $T = \Phi_\mu/\Phi_{\rm tot}$ and 
$R = \Phi_e/\Phi_{\tau}$, where 
$\Phi_{\rm tot}=\Phi_e + \Phi_\mu + \Phi_\tau$. We showed the 
dependence of these ratios on the mixing angles $\theta_{13}$ and 
$\theta_{23}$ for three kinds of sources of 
ultra high energy neutrinos -- pion, muon-damped and neutron 
sources. Their dependence on $\theta_{23}$ is  largest, and 
from analytical considerations we gave an idea of the extent to 
which the flux ratios could be used to constrain it. 

We defined a $\chi^2$ function to quantify the 
extent to which $\theta_{23}$ can be constrained by 
neutrino telescopes. We assumed a simplistic approach where
we worked with the flux ratios themselves but we are aware that 
in a realistic analysis one should work with ratios of actual 
number of events, taking into account the detector 
response and efficiencies. However, we have 
chosen to work in a simplified set-up because the purpose of 
this paper was to make a ballpark estimate of the 
sensitivity of neutrino telescopes to $\theta_{23}$. 
For the first time we included ``impure'' initial neutrino 
fluxes in a statistical 
analysis by introducing two variables, $\zeta$ and $\eta$, which 
parameterize deviations from the initial flux compositions  
$(1 : 2 : 0)$, $(0 : 1 : 0)$ or $(1 : 0 : 0)$ for the 
pion, muon-damped and neutron sources. 

We performed the statistical 
analysis using $T$ from one given source 
at a time and conclude that 
the best $\theta_{23}$ 
sensitivity comes from neutron sources. We showed that an error 
of less than 20 \% is necessary to obtain results comparable to 
oscillation experiments. 
We presented results by combining 
pion and neutron sources, muon-damped and neutron sources, and 
finally all three taken together. The bound improves mainly 
because neutron sources have a much better handle on $\theta_{23}$. 
We next combined $T$ and $R$ measurements at the neutrino telescopes. 
Adding the information of $R$ improves the 
$\theta_{23}$ bound substantially, but increases somewhat the 
impact of impure sources.  
In particular, we note that the combination of $T$ and $R$ for 
neutron sources could give us bounds on $\theta_{23}$ which 
are much better than the one we have currently. In fact, 
measuring $T$ and $R$ for neutron sources could be comparable 
to the bounds on $\theta_{23}$ we expect from future 
long baseline and atmospheric neutrino experiments,  
depending of course on the 
uncertainty on the measurement of the flux ratios at the 
neutrino telescopes. 
We performed the statistical test on $\theta_{23}$ for 
true $\sin^2\theta_{23}=0.5$ and $\sin^2\theta_{23}=0.6$. 
For the latter case we checked if it would be possible to 
establish the right octant of $\theta_{23}$.

In conclusion, in favorable but not unrealistic situations 
we can indeed expect useful and 
complementary limits on $\theta_{23}$ which are comparable to 
the ones from dedicated oscillation experiments.

\vspace{0.3cm}
\begin{center}
{\bf Acknowledgments}
\end{center}

We would like to thank J.~Kopp and M.~Lindner for helpful discussions. 
V.N.~and W.R.~were supported by the Deutsche Forschungsgemeinschaft 
in the Sonderforschungsbereich 
Transregio 27 ``Neutrinos and beyond -- Weakly interacting particles in 
Physics, Astrophysics and Cosmology''.  
S.C.~acknowledges support from the Neutrino Project
under the XI Plan of Harish-Chandra Research Institute.

\pagestyle{empty}

\thispagestyle{empty}

\end{document}